arXiv:1512.01478v2 [cs.NI] 22 Sep 2017

# A Stochastic Geometry Approach to Asynchronous Aloha Full-Duplex Networks

Andrea Munari, *Member IEEE*, Petri Mähönen, *Senior Member IEEE*, Marina Petrova, *Member IEEE*

***Abstract*—In-band full-duplex is emerging as a promising solution to enhance throughput in wireless networks. Allowing nodes to simultaneously send and receive data over the same bandwidth can potentially double the system capacity, and a good degree of maturity has been reached for physical layer design, with practical demonstrations in simple topologies. However, the true potential of full-duplex at a system level is yet to be fully understood. In this paper, we introduce an analytical framework based on stochastic geometry that captures the behaviour of large full-duplex networks implementing an asynchronous random access policy based on Aloha. Via exact expressions we discuss the key tradeoffs that characterise these systems, exploring among the rest the role of transmission duration, imperfect self-interference cancellation and fraction of full-duplex nodes in the network. We also provide protocol design principles, and our comparison with slotted systems sheds light on the performance loss induced by the lack of synchronism.**

## I. Introduction

THE inability for a radio to concurrently send and receive information over the same frequency band has represented for decades a cornerstone in the design of wireless communications systems. The conceptually simple task of cancelling a known radiated message superposed to an incoming waveform of interest has in fact long remained elusive for real-time implementations, due to the unbalance of several orders of magnitude between useful power and self-interference. This classical state of the art has witnessed a tremendous change of perspective in the past few years, driven by the compelling quest for higher spectral efficiency, and supported by steady improvements in signal processing and computational power. Resorting to advanced self-interference suppression techniques both in the analog and digital domain, in-band full-duplex was demonstrated for the first time in practical wireless scenarios in the early 2010s [1], [2], and has captured increasing attention within the research community ever since. While detailed implementation aspects are beyond the scope of this article – the interested reader is referred to [3]–[5] for excellent surveys – most of the proposed solutions have been validated via software-defined radios or through prototypes employing off-the-shelf hardware, proving the viability of full-duplex for a large market of low-cost terminals.

With this horizon in mind, information-theoretic tools helped in clarifying the potential of concurrent transmission and reception in toy topologies, exploring the achievable rate regions and the degrees of freedom in relation to the accuracy of self-interference cancellation and the number of available antennas [6]–[9]. Alongside with these efforts, a number of protocols at the medium access layer have been designed to bridge the gap with existing standards, showing by means of simulations and testbed results interesting improvements over a half-duplex configuration in centralised [10] and distributed ad hoc scenarios [11]–[14] based on IEEE 802.11.

Despite the maturity reached in the design of a single link and the remarkable performance gains achieved in simple settings, a deep understanding of the role played by full-duplex in more complex ad hoc networks is however still elusive. In fact, when instantiated in large topologies, the novel paradigm triggers a non-trivial tradeoff between spatial reuse and aggregate interference. On the one hand, the ability to leverage simultaneous bidirectional data exchanges within node pairs allows more links to be active per unit area, potentially boosting performance. On the other hand, the additional amount of interference generated by a more aggressive access to the medium besets ongoing receptions, decreasing the probability of successfully retrieving information. The overall balance of such counterposed effects as well as the influence of more realistic traffic patterns on full-duplex connections are only a few of the open questions that still need to be addressed. Particularly, a clear theoretical perspective to derive some general design principles for next-generation wireless networks is yet to be completely grasped.

A first relevant step in this direction was taken in [15], leaning on the well-known protocol model of Gupta and Kumar. Assuming that power only propagates within a circle area centred at the transmitter, [15] evaluated the throughput gain over half-duplex in linear and lattice topologies with centrally scheduled communications, and proved that the harsher interference level brought by full-duplex intrinsically prevents a network-wide throughput doubling. Further light on the achievable performance has been recently shed resorting to stochastic geometry, which lends itself excellently to characterise the tradeoff between spatial reuse and aggregate

A. Munari and P. Mähönen are with the Inst. for Networked Systems (iNETS) of RWTH University Aachen, Kackertstrasse 9, 52072 Aachen, Germany (e-mail: {andrea.munari, petri.mahonen}@inets.rwth-aachen.de).
M. Petrova is with KTH Royal Institute of Technology, School of Information and Communication Technology, Department of Communication Systems, SE-100 44 Stockholm (email: petrovam@kth.se)

interference.[1] Some preliminary insights were offered in [17], capturing the approximated behaviour of wireless full-duplex links affected by log-normal shadowing and Nakagami fading. These results were then significantly broadened by the contribution of Tong and Haenggi [18], which is particularly relevant to the present work. The paper focused on bipolar networks, where pairs of nodes distributed over the plane following a Poisson point process access the medium via a *slotted* Aloha policy. In this context, the probability distribution of the signal-to-noise-and-interference ratio experienced at a receiver subject to Rayleigh fading was derived, accounting for both half- and full-duplex transmissions within the system. The authors confirmed that a factor-two throughput improvement over a purely half-duplex configuration is not achievable in large distributed topologies, and identified the fraction of links that shall be operated in full-duplex mode to maximise performance. The role played by residual self-interference was also analytically captured, showing how the use of full-duplex becomes in fact detrimental for the aggregate network unless a minimum level of cancellation is granted. Similar trends were confirmed by [19] for a multi-tier network where different node pairs can employ distinct transmission power levels, and extended to characterise the stability of the system's queues in [20].

While pivotal, the aforementioned results assume *perfect synchronisation* among all nodes to ensure a slotted medium contention. This condition, though, is hardly met in practical instances of distributed topologies, which resort instead to asynchronous random access strategies [21]. To tackle this gap, research has recently started to focus on listen-before-talk access schemes, extending the vast body of literature on stochastic geometry models for CSMA (see, e.g. [22] and references therein) to account for full-duplex capable nodes. Along this line, [23] derived approximated performance expressions under the protocol- and a Rayleigh fading-model considering perfect self-interference cancellation. Most notably, the work showed how the larger exclusion regions induced by a bi-direction transmission between a node pair can indeed be beneficial to ease the hidden and exposed terminal problems. As a result, relevant aggregate throughput gains over a purely half-duplex counterparts were proven possible also in CSMA settings, particularly when the link distance is small compared to the carrier sense range.

Research efforts carried out so far, however, have provided no insights on the potential of the novel paradigm in asynchronous random access topologies that do not resort to listen-before-talk. On the other hand, this domain is gaining renewed interest thanks to emerging market applications for machine-type communications (MTC), which aim at connecting a massive number of low-power/low-complexity devices possibly spread over a wide area and exchanging information in a fully decentralised and sporadic fashion [24]. For these systems CSMA is often not the access strategy of choice, due to its well-known inefficiencies under low channel loads and large propagation times, as well as to the additional energy consumption it entails for sensing the medium [21], [25]. In this perspective, the simpler *unslotted* Aloha solution can be more convenient, and is in fact being implemented again in a number of commercial solutions such as LoRa [26] or SigFox [27]. At the same time, simultaneous transmission and reception capabilities are regarded with increasing interest for MTC as well [24], due to the dramatic network capacity improvements they might bring. Understanding whether full-duplex can be effectively leveraged in such settings is thus paramount for proper system design, and represents a deciding element towards identifying applications that can truly benefit from the new technology.

[1]While the focus of this paper is on ad hoc networks, the concept of full-duplex is being explored also in cellular scenarios. The interested reader is referred to, e.g. [16].

Motivated by this background, in the present article we tackle then a fundamental and still open question, offering the first study of in-band full-duplex for asynchronous random access networks that do not rely on carrier sensing. Our contributions can be summarised as follows:

- through stochastic geometry tools, we capture the behaviour of *unslotted* Aloha systems where a fraction of nodes are full-duplex capable. Exact expressions for the success probability at a link level as well as for the aggregate network throughput are derived, accounting for imperfect self-interference cancellation and clarifying the impact of design parameters such as transmission duration and distance between transmitter and receiver. In the discussion, similarities and differences with the seminal work in [18] for slotted systems are highlighted, stressing the peculiarities of asynchronous access;
- assuming all links to be of the same duration, we identify different optimal operating regions for the system. In particular, we show how full-duplex shall be preferred for short packet exchanges, whereas for transmissions longer than a specified threshold the whole network shall be operated in half-duplex to avoid throughput losses;
- we explore the additional degree of freedom of having transmissions of different duration in the network, and derive the optimal working configuration in this case, significantly extending the applicability of the analytical framework;
- we finally present a direct comparison with the results derived in [18] for slotted Aloha, and discuss the throughput degradation induced by the lack of synchronisation among nodes, offering a tool to approach a key cost-performance tradeoff in system planning.

We start our study in Section II introducing the system model and some preliminary results on stochastic geometry, later leveraged in Section III to derive the performance of an asynchronous full-duplex network. Sections IV and V extend the framework to account for different packet durations and offer a comparison with slotted schemes, eventually leading to the conclusions drawn in Section VI.

## II. System Model and Preliminaries

Throughout this paper we focus on an infinite population of users spread over the plane that share a common medium to exchange information in the form of data packets. Nodes are organised in pairs or *clusters*,[2] and only one-hop links between a terminal and its pair-companion are considered. We indicate the two users forming cluster $\mathcal{C}_i$ as $\mathsf{U}_i$ and $\mathsf{V}_i$, and specify their location by means of the two-dimensional coordinate vectors $\mathbf{u}_i$ and $\mathbf{v}_i$ respectively, as shown in Fig. 1. For each pair, we assume nodes to be separated by a common distance $r \geq 1$, i.e. $\|\mathbf{u}_i - \mathbf{v}_i\| = r, \ \forall i$.[3]

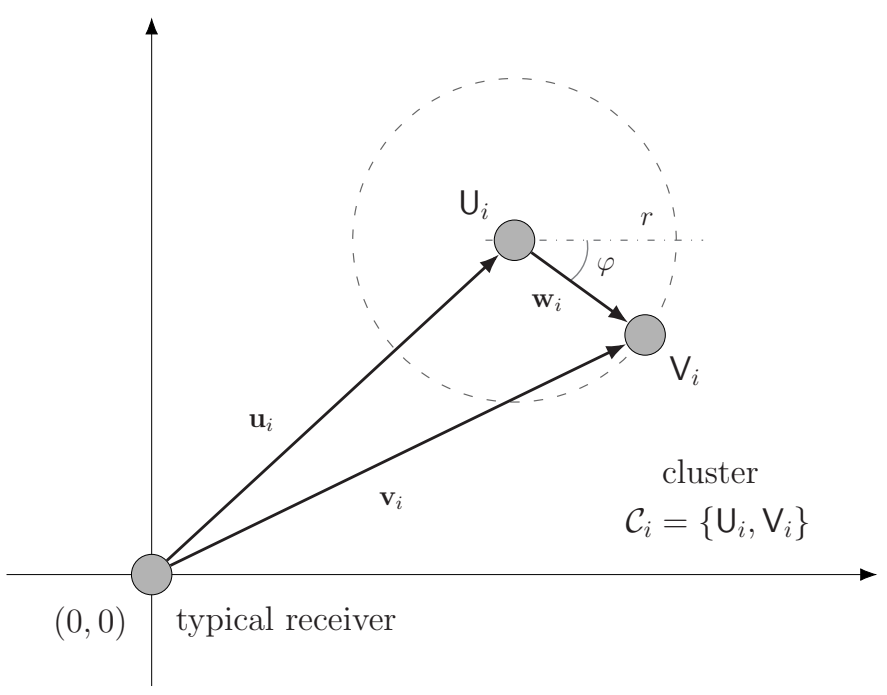

Fig. 1: Reference geometry for the links between the typical receiver located at the origin and nodes within cluster $\mathcal{C}_i$.

Clusters can operate either in *half-duplex* or *full-duplex* mode. In the former case, one terminal transmits packets according to the underlying medium access strategy, while the other simply acts as receiver and does not generate any traffic. As to the latter mode, instead, both nodes in the pair access the channel simultaneously, leveraging in-band full-duplex capabilities to send and receive data at the same time over the same frequency. In order to cover a broad range of configurations, we assume that a fraction $q$ of clusters resort to full-duplex operations, whereas the remaining perform half-duplex access.

Communication parameters in terms of bitrate, modulation and coding are, unless otherwise specified, common to all terminals and set for every transmission to have a duration $D$. Medium contention is performed asynchronously via random access, following a simple unslotted Aloha protocol, and no feedback nor retransmission policy are considered. Following this approach, the channel is accessed without sensing the surrounding activity and without adhering to any slotted time structure. The only form of coordination we consider in the network is thus between users of the same pair, assuming the receiver to be always listening to the incoming transmission of its companion in a half-duplex cluster, or both nodes to simultaneously start a reciprocal transmission in the full-duplex configuration. In this perspective, while we abstract here any further details for the sake of generality and mathematical tractability, it is worth mentioning that different strategies have been proposed in the literature to bring such coordination into practice, see, e.g. [10], [11], [13].

In order to capture the behaviour of the system, we introduce an analytical framework based on stochastic geometry, and model links in the network via a homogeneous time-space Poisson point process (PPP) $\Lambda = \{(\mathbf{u}_i, T_i)\}$ of intensity $\lambda$. With reference to the topology of Fig. 1, the spatial component $\mathbf{u}_i$ of the process describes the location of terminal $\mathsf{U}_i$, also referred to as the *cluster centre*, which, in case of a half-duplex pair, acts as transmitter. The companion node $\mathsf{V}_i$ is instead randomly scattered over a circle of radius $r$ centred at $\mathsf{U}_i$, allowing its position to be expressed as $\mathbf{v}_i = \mathbf{u}_i + \mathbf{w}_i$, where $\mathbf{w}_i = re^{j\varphi_i}$ and $\varphi_i \sim \mathcal{U}[0, 2\pi)$. In turn, the temporal component $T_i$ of $\Lambda$ identifies the start time of the transmission performed by cluster $\mathcal{C}_i$. From this standpoint, it is important to stress that we do not focus on a spatial point process describing the distribution of nodes and track the evolution of their unslotted medium access over time, e.g. by resorting to a renewal process. Conversely, $\Lambda$ jointly captures position and transmission time of a cluster, modelling the network as a population of node pairs that are born at a random time and a random location and occupy the channel for a predefined duration $D$ before disappearing. This approach introduces a further level of abstraction, as it embeds in the sole parameter $\lambda$ the spatial density of the population and the traffic generation pattern, as well as possible backoff strategies applied to the medium access contention. On the other hand, these working hypotheses will yield a compact mathematical formulation capable of identifying the key tradeoffs of full-duplex networks, and will be shown to offer a very accurate estimation of the performance of unslotted systems in Section III. Moreover, from a practical angle not only can the space-time process under consideration easily be mapped to mobile topologies, but also to scenarios with very large populations of terminals generating sporadic traffic, covering a case of strong interest for MTC applications.

Within this framework, $\Lambda$ characterises the number of links initiated in a region of area $A$ over a time interval $T$, described as a Poisson r.v. of parameter $\lambda A T$. The hybrid nature of clusters in the network is accounted for by having each pair independently decide whether to establish a half- or a full-duplex connection with probability $1-q$ and $q$, respectively. By virtue of the properties of thinning for PPPs (see, e.g. [28]), the original process can then be conveniently expressed as $\Lambda = \Lambda_{\mathsf{hd}} \cup \Lambda_{\mathsf{fd}}$, where $\Lambda_{\mathsf{hd}}$ and $\Lambda_{\mathsf{fd}}$ are two independent PPPs of intensity $(1-q)\lambda$ and $q\lambda$.

Wireless links are affected by path loss and Rayleigh fading, and the coherence time of the wireless channel is assumed long enough for fading coefficients to remain constant throughout the duration of a packet exchange. Accordingly, we model the power received by a node at position $\mathbf{y}$ from a transmission originated at location $\mathbf{x}$ as $PL(\mathbf{x}, \mathbf{y})\zeta$. Here, $P$ is the transmission power common to all users, $\zeta$ is an exponential random variable with unit mean and pdf $f_\zeta(a) = e^{-a}, a \geq 0$ describing fading, and $L(\mathbf{x}, \mathbf{y})$ accounts for the signal attenuation. With the aim of focussing on the key performance drivers, we do not explicitly address antenna gains or other propagation factors, and consider instead a simple path loss law based on the distance $d = \|\mathbf{x} - \mathbf{y}\|$, in the form $L(d) = d^{-\alpha}, \ \alpha > 2$.[4]

[2]The terms cluster and node pair will be used interchangeably.

[3]The presented results can easily be extended to an arbitrary i.i.d. distribution for the pair distance. Having a single value for $r$, however, eases the mathematical discussion while prompting all the key tradeoffs.

[4]Although rigorously the considered path loss law is only meaningful for $d \geq 1$, extensive results have shown its capability of properly capturing the behaviour of large networks, see, e.g. [28].

The effectiveness of the presented system clearly lives on the tradeoff between the spatial reuse enabled by random access and the mutual interference that besets concurrent links. From this standpoint, it is relevant to stress that the asynchronous nature of the MAC layer under consideration can induce an interference level that is not constant even within the duration of a packet exchange. Starting from this remark, and leveraging the homogeneity of the PPP $\Lambda$, we focus without loss of generality on the *typical receiver*, i.e. a node at the origin of the plane whose incoming data unit starts at time 0, and derive the time-varying interference $I(t)$ it perceives at a generic instant $t \in [0, D]$. To this aim, it is useful to resort to the indicator function $\mathbb{I}(\cdot)$ and introduce the ancillary operator $\iota(t) = \mathbb{I}(T \leq t \leq T+D)$, specifying whether cluster $(\mathbf{u}, T)$ is active at time $t$. Moreover, we simplify without risk of confusion the notation for the path loss, setting for an arbitrary point $\mathbf{x}$ on the plane $L(\mathbf{x}) := L(\mathbf{x}, \mathbf{0})$. Leaning on these steps, it is possible to isolate the interference contributions $I_{\mathsf{hd}}(t)$ and $I_{\mathsf{fd}}(t)$ of half- and full-duplex pairs, with the former having only one terminal sending data, while the latter triggering two spatially disjoint and concurrent transmissions. The time-varying interference perceived at the typical receiver can thus be expressed as $I(t) = I_{\mathsf{hd}}(t) + I_{\mathsf{fd}}(t)$, where

$$\begin{aligned} I_{\mathsf{hd}}(t) &= \sum_{(\mathbf{u},T)\in\Lambda_{\mathsf{hd}}} \iota(t)\cdot P\, L(\mathbf{u})\, \zeta_{\mathbf{u}} \\ I_{\mathsf{fd}}(t) &= \sum_{(\mathbf{u},T)\in\Lambda_{\mathsf{fd}}} \iota(t)\cdot P\,\big(\, L(\mathbf{u})\zeta_{\mathbf{u}} + L(\mathbf{v})\zeta_{\mathbf{v}}\,\big) \end{aligned} \tag{1}$$

and, in the second equation, $\zeta_{\mathbf{u}}$ and $\zeta_{\mathbf{v}}$ indicate the independent fading coefficients for the links between the receiver of interest and the two nodes in the full-duplex cluster $(\mathbf{u}, T)$. Given the interference-limited nature of the networks under consideration, we disregard thermal noise and evaluate the performance of the system based on the signal-to-interference ratio (SIR). More specifically, in an effort to preserve the mathematical tractability of the problem, we are interested in the ratio between the incoming power of the desired signal at a receiver and the time-averaged value $\mathcal{I}$ of the interference it perceives over the packet.[5] Within our framework, the latter quantity and its half- and full-duplex clusters induced components $\mathcal{I}_{\mathsf{hd}}$ and $\mathcal{I}_{\mathsf{fd}}$ can be expressed as

$$\mathcal{I} = \mathcal{I}_{\mathsf{hd}} + \mathcal{I}_{\mathsf{fd}} = \frac{1}{D}\int_0^D I_{\mathsf{hd}}(t)\, dt + \frac{1}{D}\int_0^D I_{\mathsf{fd}}(t)\, dt\,. \tag{2}$$

We thus introduce the signal-to-time-averaged-interference ratio for a half-duplex receiver as $\mathsf{SIR}_{\mathsf{hd}} = PL(r)\zeta/\mathcal{I}$. On the other hand, decoding during a full-duplex connection is also hampered by a residual self-interference component $\mathcal{S}$ due to imperfect cancellation algorithms, so that $\mathsf{SIR}_{\mathsf{fd}} = PL(r)\zeta/(\mathcal{I}+\mathcal{S})$. Buttressed by experimental results [29], we assume a linear dependence of $\mathcal{S}$ to the emitted power, and relate it to a cancellation efficiency coefficient $\eta \in [0, 1]$ as $\mathcal{S} = P(1-\eta)$. This working hypothesis is particularly handy, as it induces SIR values which are independent of $P$ for both half- and full-duplex links, helping to identify broadly applicable tradeoffs. Accordingly, in the remainder of our discussion we will refer to the case of unit transmission power, and set $P = 1$ without loss of generality.

[5]The assumption of threshold decoding based on the introduced definition of SIR is representative of, e.g. systems using coding and interleaving.

A threshold model is considered for decoding, with a packet being retrieved if the SIR experienced at its receiver is above a reference value $\theta$. Focussing first on a half-duplex cluster, the probability of its data exchange to be successful directly follows:

$$p_s^{(\mathsf{hd})} = \mathbb{P}\{\mathsf{SIR}_{\mathsf{hd}} \geq \theta\} = \mathbb{P}\left\{\zeta \geq \frac{\theta\,(\mathcal{I}_{\mathsf{hd}} + \mathcal{I}_{\mathsf{fd}})}{L(r)}\right\}. \tag{3}$$

Starting from the exponential distribution of $\zeta$ and leveraging the law of total probability over the two independent processes $\Lambda_{\mathsf{hd}}$ and $\Lambda_{\mathsf{fd}}$, (3) can be conveniently written as

$$p_s^{(\mathsf{hd})} = \mathbb{E}\left[e^{-\mathcal{I}_{\mathsf{hd}}\,\theta\, r^{\alpha}}\right]\mathbb{E}\left[e^{-\mathcal{I}_{\mathsf{fd}}\,\theta\, r^{\alpha}}\right] = \mathcal{L}_{\mathcal{I}_{\mathsf{hd}}}(\theta r^{\alpha})\,\mathcal{L}_{\mathcal{I}_{\mathsf{fd}}}(\theta r^{\alpha}) \tag{4}$$

where the second equality stems from the definition of Laplace transform of a random variable $X$, $\mathcal{L}_X(s) := \mathbb{E}[e^{-sX}]$. Following a similar approach, it is straightforward to also derive the success probability for a full-duplex link as

$$p_s^{(\mathsf{fd})} = e^{-(1-\eta)\theta r^{\alpha}}\cdot \mathcal{L}_{\mathcal{I}_{\mathsf{hd}}}(\theta r^{\alpha})\,\mathcal{L}_{\mathcal{I}_{\mathsf{fd}}}(\theta r^{\alpha}) = \beta\, p_s^{(\mathsf{hd})} \tag{5}$$

where $\beta := \exp\big(-(1-\eta)\theta r^{\alpha}\big)$ accounts for imperfect self-interference cancellation. It is worth noting that the formulations of (4), (5) resemble the ones in [18]. The factorisation in terms of Laplace transforms, in fact, stems from the independent thinning of $\Lambda$ in its half- and full-duplex components in combination with the underlying random access policy. On the other hand, the way interference affects ongoing links, i.e. the specific expression of $\mathcal{L}_{\mathcal{I}_{\mathsf{hd}}}$ and $\mathcal{L}_{\mathcal{I}_{\mathsf{fd}}}$, is profoundly different in the asynchronous case under study and the slotted Aloha setting of [18], as will become clear in the next section.

In order to complement our analysis, we evaluate the performance of the system in terms of the throughput density $\mathcal{T}$, defined as the average number of information bits per second successfully exchanged in the network per unit area. Assuming an information bitrate of $W$ bit/s common to all transmissions, a delivered packet contributes with $WD$ bits to the throughput, so that

$$\mathcal{T} = \lambda D\, W\Big((1-q)p_s^{(\mathsf{hd})} + 2q\, p_s^{(\mathsf{fd})}\Big). \tag{6}$$

Within (6), the first addend accounts for the fraction $1-q$ of half-duplex connections, whereas the second brings in the contribution of full-duplex clusters, potentially delivering up to two data units per link.

## III. The Performance of Asynchronous Full-Duplex networks

The model of Section II highlighted how the performance of our asynchronous system can be characterised through the Laplace transforms $\mathcal{L}_{\mathcal{I}_{\mathsf{hd}}}(s)$ and $\mathcal{L}_{\mathcal{I}_{\mathsf{fd}}}(s)$. An elegant formulation of the former was devised by Błaszczyszyn et al. in the context of *solely half-duplex* Aloha networks [30]. For the sake of compactness we omit here the details of their derivation, and rather focus on a slightly modified version of the original outcome [30, (3.8)] obtained via simple mathematical manipulations. Accordingly, we express the Laplace transform of the interference perceived at the typical receiver due to half-duplex pairs in the form

$$\mathcal{L}_{\mathcal{I}_{\mathsf{hd}}}\left(\theta r^{\alpha}\right) = \exp\Big(-\lambda(1-q)\, D\, \Omega_{\mathsf{hd}}\Big) \tag{7}$$

where the ancillary function $\Omega_{\mathsf{hd}}$ is defined as

$$\Omega_{\mathsf{hd}}(r,\theta,\alpha) = \pi r^2\, \theta^{\frac{2}{\alpha}}\, \Gamma\left(1+\frac{2}{\alpha}\right)\Gamma\left(1-\frac{2}{\alpha}\right)\frac{2\alpha}{(\alpha+2)} \tag{8}$$

and $\Gamma(x) = \int_0^\infty x^{t-1}e^{-x}\,dt$ is the complete Gamma function. The result in (7) is particularly insightful, as it isolates the role of two key performance drivers. On the one hand, an exponential dependence of $\mathcal{L}_{\mathcal{I}_{\mathsf{hd}}}$ – and thus of the success probability – on the duration $D$ of the active links is prompted, stressing the intrinsic weakness of longer transmissions to interference. On the other hand, the factor $\Omega_{\mathsf{hd}}$ summarises the impact of the system parameters $r$, $\theta$ and $\alpha$, embedding the structure of the interference generated by half-duplex clusters.

### *A. The Laplace transform of interference by full-duplex links*

A finer level of detail is needed instead to determine the impact of full-duplex pairs accounted for by $\mathcal{L}_{\mathcal{I}_{\mathsf{fd}}}(\theta r^{\alpha})$, which represents a fundamental and novel contribution of our framework. As a preliminary step, the value of the generated interference $\mathcal{I}_{\mathsf{fd}}$ introduced in (2) can be conveniently simplified by recalling the definition in (1) to obtain

$$\mathcal{I}_{\mathsf{fd}} = \sum_{(\mathbf{u},T)\in\Lambda_{\mathsf{fd}}} \omega(T)\cdot\big(L(\mathbf{u})\,\zeta_{\mathbf{u}} + L(\mathbf{v})\,\zeta_{\mathbf{v}}\big)$$

where the time averaging is captured by the function

$$\omega(T) \triangleq \int_0^D \frac{\iota(t)}{D}\,dt = \begin{cases} \frac{D-|T|}{D} & T\in[-D,D] \\ 0 & \text{elsewhere} \end{cases}.$$

Taking the lead from this, we can set the calculation of the the Laplace transform $\mathcal{L}_{\mathcal{I}_{\mathsf{fd}}}(s)$, $s\in\mathbb{R}^+$, in the form:

$$\mathcal{L}_{\mathcal{I}_{\mathsf{fd}}}(s) = \mathbb{E}\left[e^{-s\mathcal{I}_{\mathsf{fd}}}\right] = \mathbb{E}\left[\prod_{(\mathbf{u},T)\in\Lambda_{\mathsf{fd}}} e^{-s\,\omega(T)(L(\mathbf{u})\zeta_{\mathbf{u}}+L(\mathbf{v})\zeta_{\mathbf{v}})}\right]. \tag{9}$$

The expectation in (9) operates over both fading and the PPP in its space and time components. As to fading, the independence of the involved Rayleigh channel coefficients allows to bring the expectation over random variables $\zeta_{\mathbf{u}}$ and $\zeta_{\mathbf{v}}$ inside the product, enabling the reformulation reported in (10) at the bottom of next page. With reference to this, equality (a) simply follows by the law of the unconscious statistician and by the exponential distribution of unit mean for the fading coefficients. On the other hand, step (b) stems from the probability generating functional of the homogeneous PPP $\Lambda_{\mathsf{fd}}$ of intensity $\lambda q$ [28], recalling that, for cluster $\mathcal{C}_i$, $\mathsf{V}_i$ is uniformly distributed around the centre node $\mathsf{U}_i$ over a circle of radius $r$, i.e. $\mathbf{v}_i = \mathbf{u}_i + re^{j\varphi}$, $\varphi\sim\mathcal{U}[0,2\pi)$ (see Fig. 1). The formulation in (10) relates the Laplace transform to the averaging over the space and time components of a full-duplex cluster, and can be further simplified under the considered working assumptions. In particular, the linear trend and limited support of $\omega(T)$ lead via straightforward calculations to a closed-form expression for the inner integration:

$$\int_{\mathbb{R}}\left(1-\frac{1}{2\pi}\int_0^{2\pi}\frac{1}{1+sL(\mathbf{u})\omega(T)}\,\frac{1}{1+sL(\mathbf{u}+re^{j\varphi})\omega(T)}\,d\varphi\right)dT$$
$$= 2D\int_0^{2\pi} 1-\frac{1}{2\pi}\,\frac{\ln\big(1+sL(\mathbf{u})\big)-\ln\big(1+sL(\mathbf{u}+re^{j\varphi})\big)}{s\big(L(\mathbf{u})-L(\mathbf{u}+re^{j\varphi})\big)}\,d\varphi \tag{11}$$

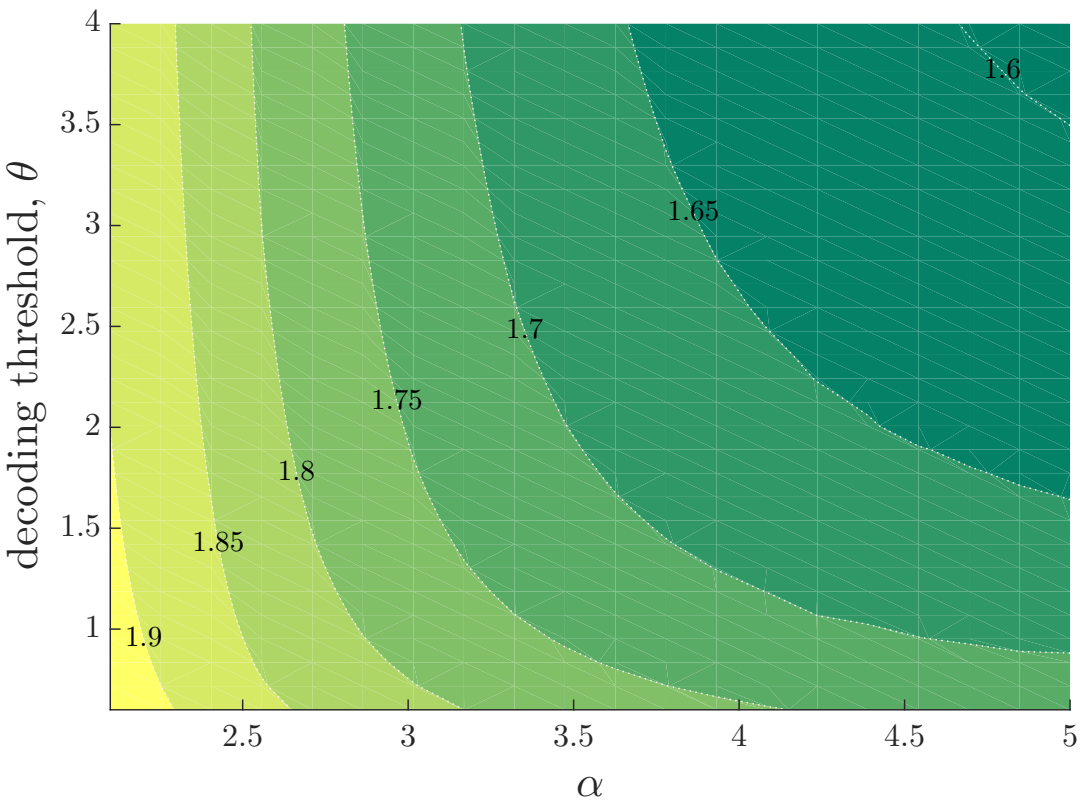

Fig. 2: Values of $\delta(\theta,\alpha) = \Omega_{\mathsf{fd}}(r,\theta,\alpha)/\Omega_{\mathsf{hd}}(r,\theta,\alpha)$ as a function of the path loss exponent $\alpha$ and of the decoding threshold $\theta$.

Plugging (11) into (10), the remaining spatial averaging can be expressed resorting to polar coordinates and observing that the overall integrand does not depend on the azimuthal component of the cluster centre due to symmetry, to obtain

$$\mathcal{L}_{\mathcal{I}_{\mathsf{fd}}}\left(\theta r^{\alpha}\right) = \exp\Big(-\lambda q\, D\, \Omega_{\mathsf{fd}}\Big) \tag{12}$$

where $\Omega_{\mathsf{fd}}$ is reported by (13) at the bottom of next page. The presented result is remarkable, as we can once more isolate the effect of the key design parameters $q$ and $D$ from the factor $\Omega_{\mathsf{fd}}$. The latter, in turn, is not affected by the fraction of full-duplex clusters in the network nor by the duration of the transmissions, and just characterises the interference contribution that each full-duplex link produces. Further insights are offered by the following, proven in Appendix A:

*Theorem 1:* Within the considered system model, for any $\alpha > 2$ and $r \geq 1$, the ratio $\Omega_{\mathsf{fd}}(r,\theta,\alpha)/\Omega_{\mathsf{hd}}(r,\theta,\alpha)$ is independent of the distance $r$ between the nodes in a cluster.

The statement entails some relevant remarks. Firstly, it confirms a similar trend exhibited in slotted systems [18, Cor. 5], prompting a non-trivial parallel given the deeply different structure of the interference in the two cases.[6] Secondly, recalling the expression of $\Omega_{\mathsf{hd}}$ in (8), we infer that $\mathcal{L}_{\mathcal{I}_{\mathsf{fd}}}$ exhibits a quadratic dependence on $r$ as well. Thus, the success probability (4) of a half-duplex data exchange is in the form of an exponential function of $r^2$ regardless of the path loss exponent $\alpha$. Even more interestingly, for $\beta = 1$ the same trend holds for $p_s^{(\mathsf{fd})}$, leading to the conclusion that the performance ratio between half- and full-duplex transmissions is not affected by the link distance under the assumption of *perfect* self-interference cancellation. In asynchronous systems, this is a peculiar trend of Aloha-based access that contrasts with the behaviour of CSMA, for which the listen-before-talk mechanism was shown to favour full-duplex gains in short links even with $\beta = 1$ [23]. As we will see later, the source-destination distance starts playing a critical role for Aloha only in the presence of residual self-interference.

[6] An extensive comparison with slotted schemes will presented in Section V.

From a different angle, Theorem 1 allows the factorisation $\Omega_{\mathsf{fd}}(r,\theta,\alpha) = \delta(\theta,\alpha)\cdot\Omega_{\mathsf{hd}}(r,\theta,\alpha)$. This expression is particularly handy, prompting the computation of $\Omega_{\mathsf{fd}}$ as the product of $\Omega_{\mathsf{hd}}$, for which a simple closed form expression is available, and a function $\delta$ which can readily be evaluated numerically. Furthermore, $\delta$ only depends on the decoding threshold and on the path loss exponent, so that it is sufficient to compute it once for a $(\theta,\alpha)$ pair to readily get the success probabilities for any value of density $\lambda$, for any fraction of full-duplex clusters as well as for any link distance. The introduced framework thus offers a compact tool to easily characterise the performance of a variety of network configurations. Along this line of reasoning, Fig. 2 reports in contour form the values of $\delta(\theta,\alpha)$ for an extensive set of parameters. In order to get a deeper understanding of the plot, it is insightful to consider the two opposite scenarios of a purely half-duplex, i.e. $q = 0$, and a purely full-duplex, i.e. $q = 1$, network. Recalling (4), the success probability for a data exchange in the former case simplifies to $p_s^{(\mathsf{hd})} = \exp(-\lambda D\Omega_{\mathsf{hd}})$. Similarly, when $q = 1$, we get from (5) and under the hypothesis of ideal self-interference cancellation $p_s^{(\mathsf{fd})} = \exp(-\lambda D\Omega_{\mathsf{fd}})$. Assuming the same configuration in terms of traffic intensity $\lambda$ and packet duration $D$, thus, the reliability loss undergone in a solely full-duplex system due to the higher interference triggered by concurrent communications between nodes of a cluster is driven exactly by the ratio $\delta$. From this standpoint, Fig. 2 quantifies two interesting trends. On the one hand, larger path loss exponents are especially beneficial to full-duplex communications, by virtue of the stronger attenuation undergone by the aggregate network interference. On the other hand, for a given value of $\alpha$, the same reduction in terms of $\theta$, i.e. the same improvement in decoding schemes to tolerate lower SIR, pays off more in a completely half-duplex configuration. In purely full-duplex settings, the beneficial effect of more advanced receivers is counteracted by the additional interference that has to be faced.

### B. Aggregate network throughput

The exact expressions derived for the packet retrieval probabilities allow us to evaluate the network throughput density, which captures the tradeoff between spatial reuse and additional interference triggered by in-band full-duplex. In particular, the general formulation of (6) can be further elaborated to obtain

$$\mathcal{T} = W\lambda D\big(1+q\,(2\beta-1)\big)\, e^{-\lambda D\big((1-q)\Omega_{\mathsf{hd}}+q\Omega_{\mathsf{fd}}\big)}. \tag{14}$$

Within (14), we can conveniently introduce the network load $\mathsf{G} = \lambda D$, quantifying the average fraction of time the medium is occupied over a reference unit area. The definition prompts how, despite the non-trivial nature of the system under study, the throughput exhibits the trend of a typical Aloha-based MAC, mathematically described by the product of $\mathsf{G}$ to a negative exponential function of $\mathsf{G}$ itself. Throughout our analysis, however, we do not characterise performance in terms of network load, but rather investigate the behaviour of the network under a fixed traffic density $\lambda$ and varying the duration of $D$ of a data exchange. The rationale behind the choice is twofold. In the first place, this approach will offer insightful hints on how to optimise systems in which topology and traffic pattern cannot be controlled. A relevant example are MTC networks, where each node of a vast population only sporadically generates a packet, so that identifying the proper length of an information unit for a half- or full-duplex exchange can become of critical relevance. Secondly, capturing the dependence of the throughput on $D$ will pave the road to the investigation of an additional degree of freedom typically missing in slotted systems, i.e. the transmission of packets of different durations.

We begin our discussion assuming ideal self-interference cancellation (i.e. $\eta = 1$), and focus on the cases of a purely half-duplex and a purely full-duplex network. Unless otherwise specified, we refer to a system with parameters $\lambda = 0.05$, $r = 1$, $\alpha = 4$, $W = 1$ and $\theta = 2$. The analytically derived throughput density achievable in the two cases when varying the packet duration $D$ is reported in Fig. 3 by solid lines. To validate the underlying rain-Poisson model, we ran dedicated simulations considering a network where the spatial and temporal component of medium access are decoupled. Specifically, node pairs were located on the plane via a PPP $\Lambda' \subset \mathbb{R}^2$ of intensity $\lambda'$. Following the pure Aloha

$$\mathcal{L}_{\mathcal{I}_{\mathsf{fd}}}(s) = \mathbb{E}_{\Lambda_{\mathsf{fd}}}\left[\prod_{(\mathbf{u},T)\in\Lambda_{\mathsf{fd}}} \mathbb{E}_{\zeta_{\mathbf{u}}}\left[e^{-s\,\omega(T)L(\mathbf{u})\zeta_{\mathbf{u}}}\right]\mathbb{E}_{\zeta_{\mathbf{v}}}\left[e^{-s\,\omega(T)L(\mathbf{v})\zeta_{\mathbf{v}}}\right]\right] \overset{(a)}{=} \mathbb{E}_{\Lambda_{\mathsf{fd}}}\left[\prod_{(\mathbf{u},T)\in\Lambda_{\mathsf{fd}}} \frac{1}{1+sL(\mathbf{u})\omega(T)}\cdot\frac{1}{1+sL(\mathbf{v})\omega(T)}\right]$$

$$\overset{(b)}{=} \exp\left(-\lambda q\int_{\mathbb{R}^2} d\mathbf{u}\int_{\mathbb{R}} dT\left(1-\frac{1}{2\pi}\int_0^{2\pi}\frac{1}{1+sL(\mathbf{u})\omega(T)}\,\frac{1}{1+sL(\mathbf{u}+re^{j\varphi})\omega(T)}\,d\varphi\right)\right) \tag{10}$$

$$\Omega_{\mathsf{fd}}(r,\theta,\alpha) = \int_0^\infty 4u\left(\pi-\int_0^\pi \frac{\ln\left(1+\theta r^\alpha u^{-\alpha}\right)-\ln\left(1+\theta r^\alpha\left(u^2+r^2+2ru\cos\varphi\right)^{-\frac{\alpha}{2}}\right)}{\theta r^\alpha\left(u^{-\alpha}-\left(u^2+r^2+2ru\cos\varphi\right)^{-\frac{\alpha}{2}}\right)}\,d\varphi\right)du \tag{13}$$

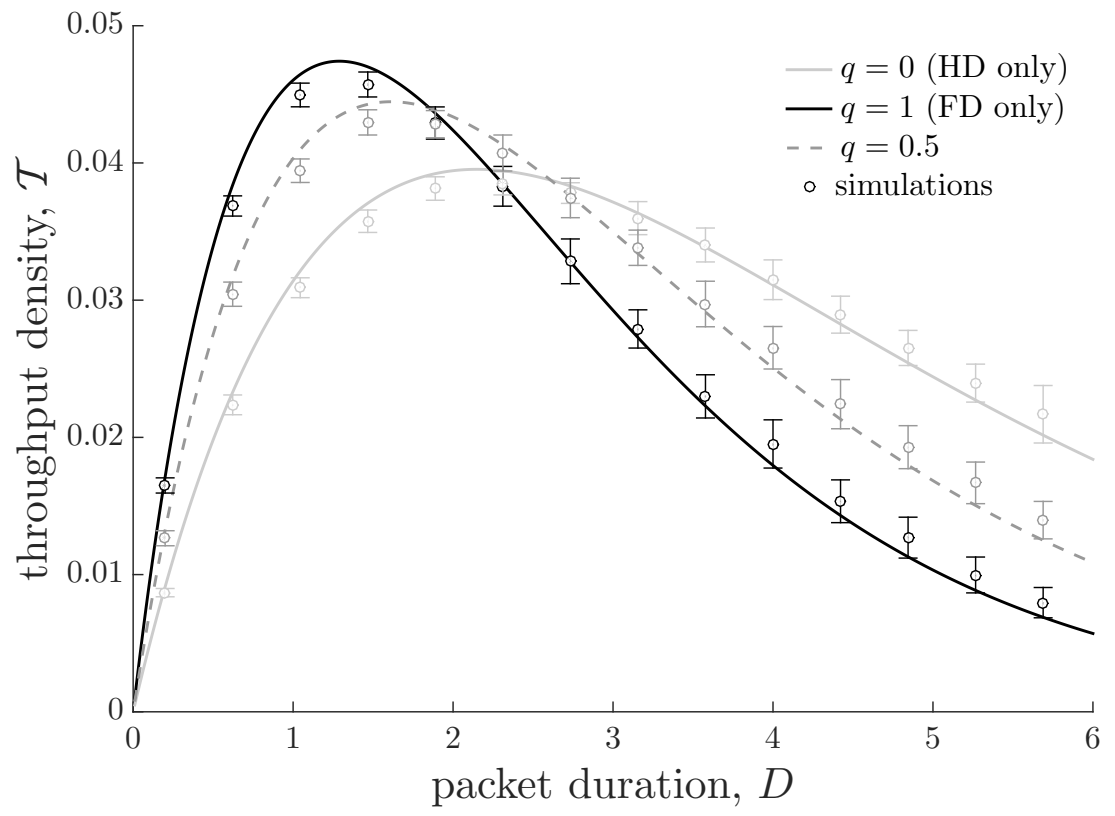

Fig. 3: Network throughput density vs packet duration for different fractions of full-duplex clusters. Lines indicate the analytical trends, whereas circled markers the results of Monte Carlo simulations.

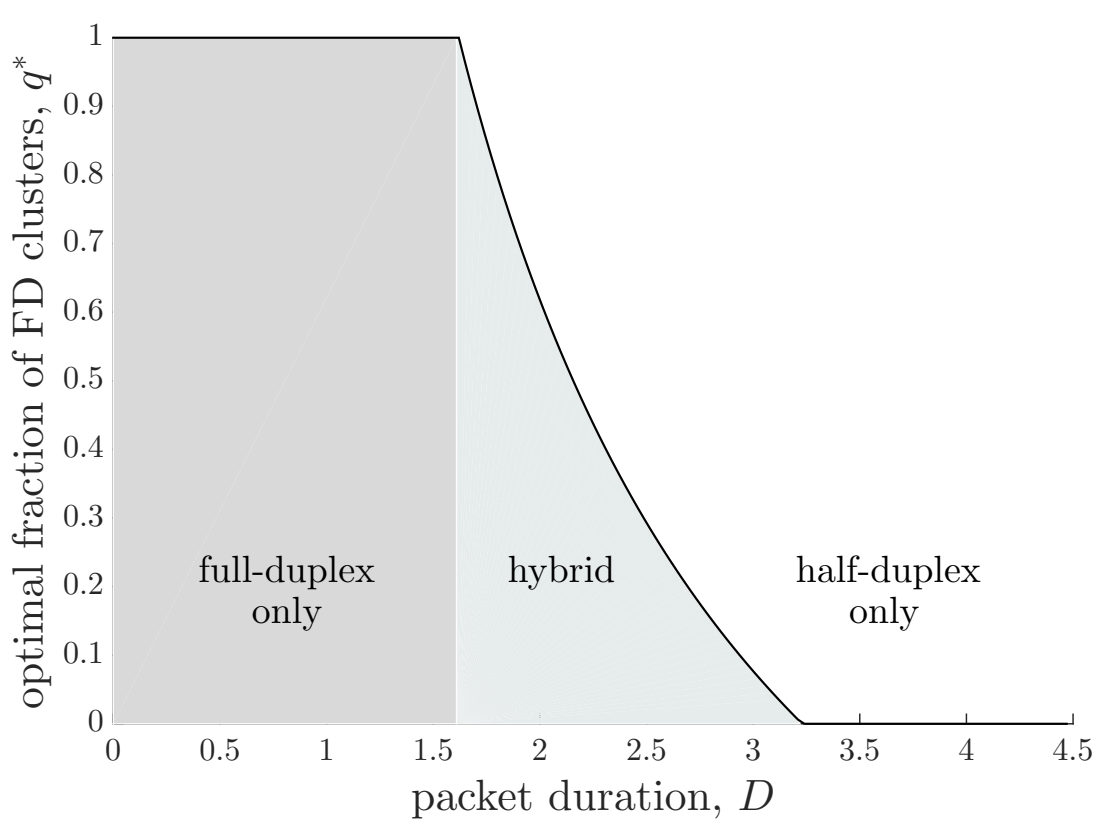

Fig. 4: Fraction of full-duplex clusters to maximise network throughput vs. packet duration (perfect self-interference cancellation).

approach, after a transmission of duration $D$ a cluster picks a random backoff uniformly drawn in $[0, B]$ and then accesses the medium again. To ensure a fair comparison, the spatial density of $\Lambda'$ was set so to have the same average network load generated by the rain Poisson process $\Lambda$, i.e. imposing $\lambda D = \lambda' D/(D+B)$. The outcome of the simulations are reported with their statistical confidence bars in Fig. 3 for $B = 14$. An excellent match with the analytical curves is shown for all values of $D$, buttressing the accuracy of the proposed framework relying on the rain Poisson assumption.

Focussing then on the performance trends, the plot highlights how full-duplex capabilities indeed boost performance for short data units, thanks to the higher degree of spatial reuse they enable. On the other hand, when longer packets are considered, the detrimental effect of the additional interference generated by having two concurrent transmissions per cluster kicks in, leading to a steep decrease in the achievable throughput and eventually making a simple half-duplex setting more convenient. Leveraging the broad applicability of (14), Fig. 3 also depicts the behaviour of a hybrid system in which only half the pairs are capable of transmitting and receiving at the same time ($q = 0.5$, dashed line). As expected, such a configuration blends benefits and drawbacks identified for the $q = 0$ and $q = 1$ cases, improving over a purely half-duplex system for short communications and degrading more gently than the full-duplex only scheme for longer information units. More interestingly, the curves also prompt a third region in which an intermediate configuration outperforms in fact both its counterparts. This observation raises then the relevant question of what is the fraction $q^*$ of full-duplex clusters one should aim for to maximise throughput given a certain value of $D$. On the one hand, if we naturally interpret $q$ as the penetration level of more advanced terminals in a traditional half-duplex network, the optimisation problem can be seen as a driver in the decision on whether to undergo the costs to further upgrade an existing system. On the other hand, even when the capabilities of deployed nodes cannot be changed, $q$ may still represent a key design parameter to tweak the fraction of links that shall in fact resort to full-duplex to leverage the non-trivial tradeoff between spatial reuse and additional interference at its utmost. The question can be effectively addressed leaning on the simple structure of (14), and recalling that $\Omega_{\mathsf{hd}}$ and $\Omega_{\mathsf{fd}}$ do not depend on $q$. Setting then $\partial\mathcal{T}/\partial q = 0$, we obtain the optimal fraction of full-duplex clusters when varying $D$:

$$q^* = \begin{cases} 1 & D \in [0, D_1) \\ \dfrac{1}{\lambda(\Omega_{\mathsf{fd}} - \Omega_{\mathsf{hd}})} - \dfrac{1}{2\beta - 1} & D \in [D_1, D_2) \\ 0 & D \geq D_2 \end{cases} \tag{15}$$

where the switching boundaries are defined as

$$D_1 = \frac{2\beta - 1}{\lambda(\Omega_{\mathsf{fd}} - \Omega_{\mathsf{hd}})} \cdot \left(1 - \frac{1}{2\beta}\right), \quad D_2 = \frac{2\beta - 1}{\lambda(\Omega_{\mathsf{fd}} - \Omega_{\mathsf{hd}})} \tag{16}$$

Confirming the intuition prompted by Fig. 3, (15) identifies three optimal operating regions, reported graphically in Fig. 4 under the assumption of perfect self-interference cancellation. For sufficiently short packets ($D \leq D_1$), network throughput is indeed maximised by letting as many clusters as possible – ideally, all of them – operate in full-duplex mode. Conversely, when data units are longer than threshold $D_2$, even full-duplex capable nodes shall not take advantage of simultaneous transmission and reception in favour of unidirectional links. Remarkably, a closed form expression is available to also characterise the optimal value $q^*$ in the intermediate region as a function of the ancillary functions $\Omega_{\mathsf{fd}}$ and $\Omega_{\mathsf{hd}}$. The plot thus presents a first design takeaway, suggesting the use of in-band full-duplex for quick information exchanges between two nodes rather than for longer connections. Such traffic patterns in turn are of strong interest, being well matched by an increasing number of applications that embody the small-data paradigm [24], and representing a core aspect in the domain of machine-type and device-to-device communications which typically generate short and sporadic data units. Even more relevant is to stress that the optimal operating regions introduced in Fig. 4 describe the network behaviour assuming terminals capable of ideally removing any trace of self-interference. The performance deterioration undergone by full-duplex when exchanging long data units, thus, cannot be eased resorting to more advanced signal processing, but rather represents an intrinsic and fundamental limitation faced by this technology in large and asynchronous networks.

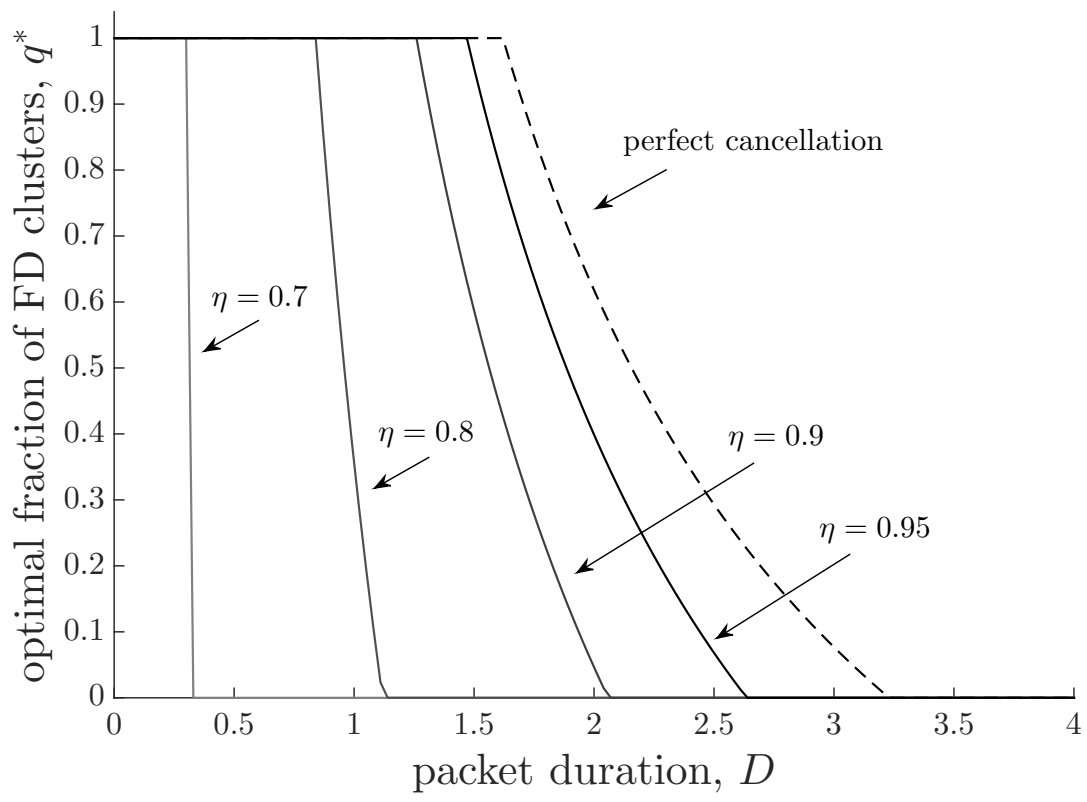

Fig. 5: Optimal fraction of full-duplex clusters to maximise network throughput vs. packet duration. Imperfect self-interference cancellation is considered.

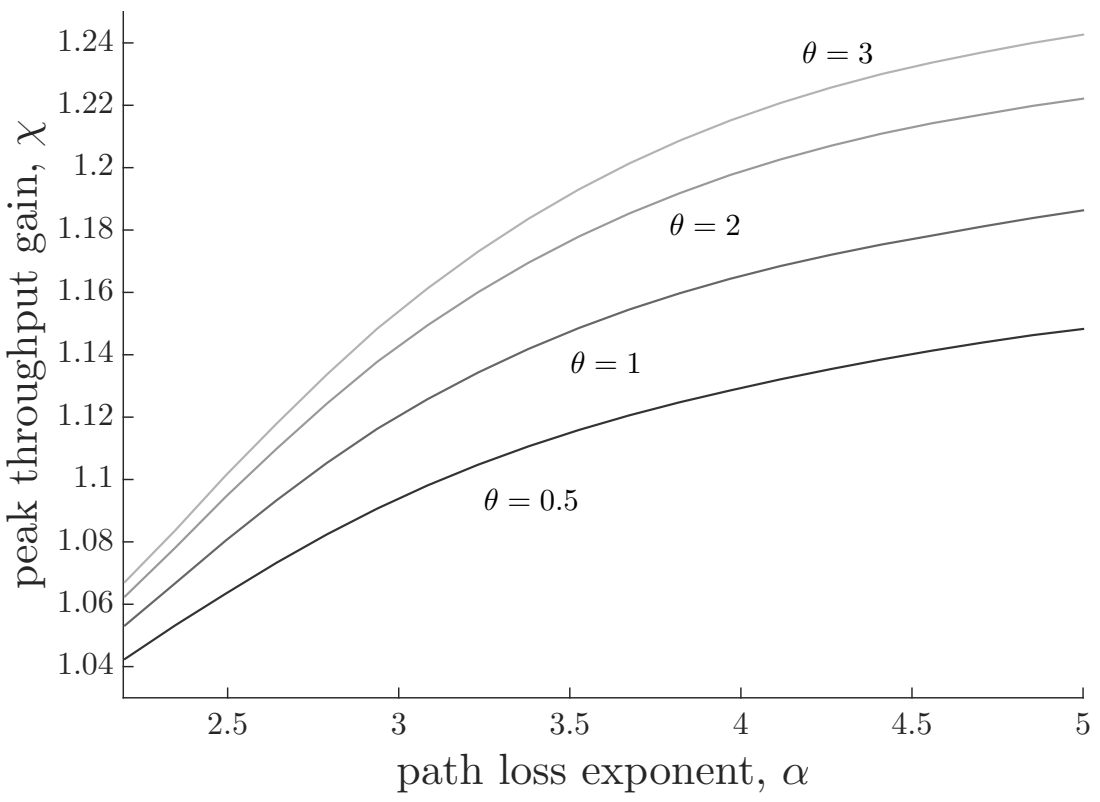

Fig. 6: Peak throughput gain achievable by a purely full-duplex network over a solely half-duplex one vs path loss exponent. Perfect self-interference cancellation is assumed.

In parallel to this general bound, the optimisation problem solved in (15)-(16) also offers a deeper understanding of the traffic patterns suitable for full-duplex in practical implementations, encompassing the impact of imperfect interference cancellation via the factor $\beta$. The outcome of the study is reported in Fig. 5, where the dashed line reproduces for completeness the regions under ideal cancellation discussed so far, whereas solid lines depict the $q^*$ against $D$ curve for different values of the efficiency parameter $\eta$. A critical role for residual self-interference decidedly emerges from the plot, as lower values of $\eta$ progressively limit the region of convenience for solely full-duplex systems to shorter communications. This trend eventually leads, for $\beta \geq 1/2$, to a situation in which a simpler half-duplex network offers better performance regardless of the packet duration. Recalling the definition of $\beta$, this translates into a minimum requirement in terms of cancellation efficiency for full-duplex to be useful in a distributed asynchronous network in the form $\eta \geq 1 - \ln(2)\, r^{-\alpha}/\theta$. As a second remark, we notice that the presence of residual self-interference induces sharper transitions between the regions where only full- and only half-duplex are to be preferred. Such a trend is in general not desirable, as the operating condition of most interest is exactly the one where both kinds of link coexist. From this standpoint, not only can intermediate values of $q$ be interpreted as representative of networks where a portion of the terminals have full-duplex capabilities, but also of topologies in which nodes within a cluster do not always have traffic for each other, failing the fundamental condition for a bidirectional connection to be established in the first place.

### C. On the maximum achievable full-duplex gain

The analysis carried out so far has clarified the importance of carefully selecting how many full-duplex links to trigger. We now extend our study to the complementary task of identifying how to tune the duration of data exchanges given a certain network configuration in terms of $q$ so to maximise performance. The solution to this problem follows from the simple dependence of $\mathcal{T}$ on the packet duration $D$ reported in (14). Straightforward calculations allow to derive the optimal operating point $D^* = 1/\left(\lambda\left((1-q)\,\Omega_{\mathsf{hd}} + q\,\Omega_{\mathsf{fd}}\right)\right)$,[7] and the corresponding throughput density as

$$\mathcal{T}^* = \frac{W\big(1+q(2\beta-1)\big)}{e\big((1-q)\,\Omega_{\mathsf{hd}} + q\,\Omega_{\mathsf{fd}}\big)}. \tag{17}$$

Remarkably, the expression does not depend on the efficiency of self-interference cancellation and on $\lambda$, showing how the peak performance is intrinsically limited by the nature of the interference generated by full-duplex links and does not scale with the density of the population. Leaning on (17), the maximum gain that bidirectional links can award when articulated topologies are considered can be computed. To this aim, we introduce the ratio $\chi$ of the peak throughput of a solely full-duplex network to the same quantity for a network operated in half-duplex mode, obtaining[8]

$$\chi := \frac{\max\big\{\mathcal{T}\,|\,q=1\big\}}{\max\big\{\mathcal{T}\,|\,q=0\big\}} = 2\beta \cdot \frac{\Omega_{\mathsf{hd}}}{\Omega_{\mathsf{fd}}}\,.$$

The metric is conveniently expressed as twice the correction factor $\Omega_{\mathsf{hd}}/\Omega_{\mathsf{fd}} = 1/\delta$, which is lower than one even under the assumption of ideal self-interference cancellation. Not only does this confirms that full-duplex can in fact not double the network capacity in large asynchronous systems, but also readily quantifies the obtainable improvement. In fact, if we initially consider the ideal case $\beta = 1$, $\chi$ is completely defined by the ancillary function $\delta(\theta, \alpha)$ introduced and already discussed in Fig. 2. For the sake of readability, the behaviour of $\chi$ against $\alpha$ for different values of the decoding threshold $\theta$ is also explicitly reported in Fig. 6. Recalling the outcome of

[7]Note that the maximisation could also be carried out over the network load, leading to a scaled result in the form $\mathsf{G}^* = \lambda D^*$. The outcomes of the following discussion thus also directly apply to load optimisation.

[8]Not surprisingly, the metric has the same form that was obtained in [18] for slotted Aloha. Indeed, the structure of $\chi$ eventually derives from the factorisation of the success probability in terms of the Laplace transforms $\mathcal{L}_{\mathcal{I}_{\mathsf{hd}}}$ and $\mathcal{L}_{\mathcal{I}_{\mathsf{fd}}}$, which, as discussed, holds in both settings. The intrinsic distinction emerges from the actual expressions of the functions that capture the different impact of the interference for synchronous and asynchronous MAC policies (see Section V).

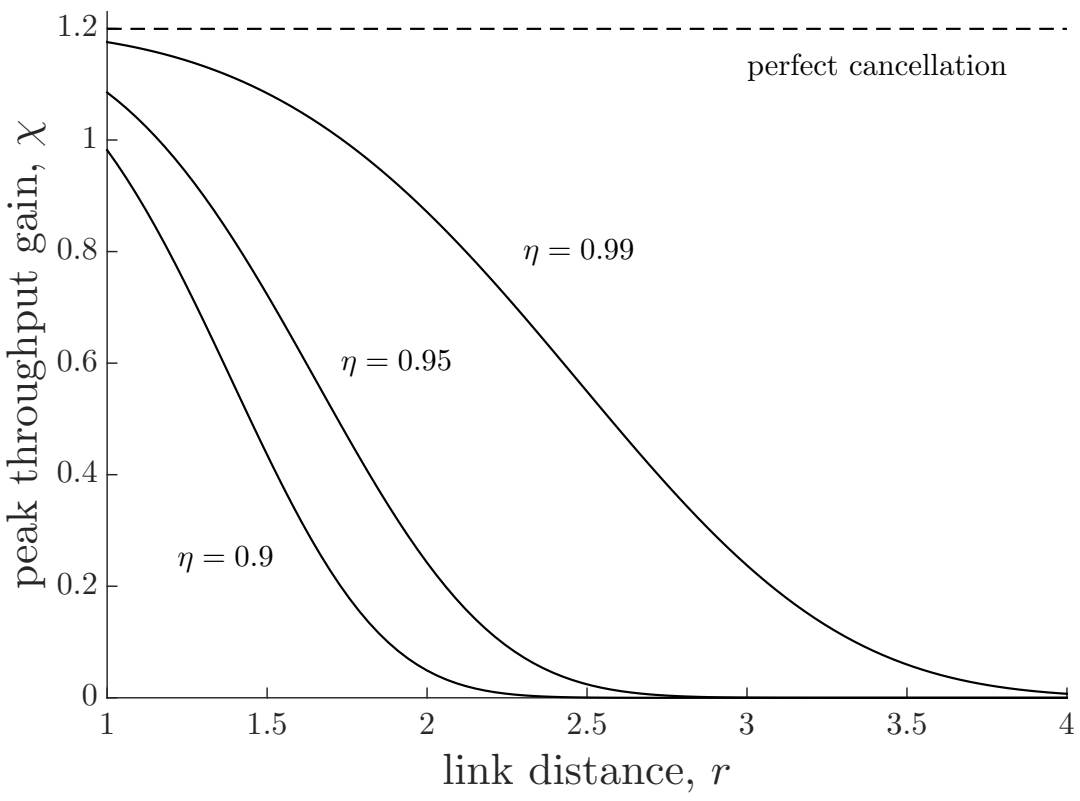

Fig. 7: Peak throughput gain achievable by a purely full-duplex network over a solely half-duplex one vs link distance.

Theorem 1 we can immediately infer that, with perfect self-interference cancellation, $\chi$ is independent of the link distance $r$. The results being presented are thus broadly applicable and characterise the fundamental behaviour of full-duplex systems, whose increased spatial reuse can pay off with at most a 20% throughput gain over their half-duplex counterparts for the reference parameter set $\alpha = 4, \theta = 2$. On the other hand, the situation drastically changes when imperfect cancellation is brought into the picture. By virtue of the exponential dependence of $\beta$ on the cluster radius, indeed, residual self-interference induces a dramatic degradation of the throughput gain offered by spatial reuse. This aspect is highlighted in Fig. 7, which reports $\chi$ as a function of $r$, and shows how already small losses in $\eta$ fundamentally limit the throughput of the full-duplex network. Even more interestingly, poorer interference cancellation levels (e.g. $\eta \leq 0.9$ in our case) eventually lead to a condition in which a purely half-duplex network outperforms its full-duplex counterpart regardless of the proximity of the communicating nodes. This offers two relevant design take-aways. In the first place, not only shall full-duplex links be employed when short data units have to be exchanged, but also they shall carefully be triggered when source and addressee are sufficiently close to each other. Secondly, the potential improvement in terms of capacity shall not distract from the importance of achieving levels of self-interference cancellation even stronger than what is desirable from an isolated-link viewpoint. In the quest for low-cost terminals, this may in fact constitute a crucial challenge.

## IV. The Impact of different Packet Durations

The framework developed in Section III extensively characterises the performance of an Aloha network with full-duplex capabilities when all transmissions occupy the channel for the same time. On the other hand, one of the key outcomes of the study has been exactly to stress how half- and full-duplex links exhibit quite distinct requirements in terms of packet duration to operate optimally, with the former supporting longer communications and the latter leveraging spatial reuse at its utmost when exchanging short information units. Such a remark triggers the natural question of whether and how the system may benefit from independently tuning the duration of links of different nature. Notably, not only would the answer pave the road for additional optimisations, but also it would shed further light on the differences between asynchronous full-duplex systems and their slotted counterparts, clarifying the role of the additional degree of freedom represented by variable packet durations for the former family.

We tackle the problem focussing on networks where each half-duplex connection has duration $D$, while a full-duplex exchange occupies the medium for $\gamma D$. For consistency, we evaluate performance in terms of the throughput density $\mathcal{T}$ introduced in (6), which readily extends in this case to

$$\mathcal{T} = \lambda W D \Big( (1-q) p_s^{(\mathsf{hd})} + 2\gamma q p_s^{(\mathsf{fd})} \Big). \tag{18}$$

The key task is thus once again to study the decoding probability for the different types of data exchanges. From this standpoint, even though the computational approach followed in Section II still holds and allows to express the success rate as the product of two Laplace transforms, the new setting slightly modifies the meaning of the involved factors. If we focus on a bidirectional connection, reception of incoming data is in the first place hampered by concurrent transmissions of clusters also operating in full-duplex mode with the same packet time. The impact of this interference contribution on the success probability is exactly the one captured by the Laplace transform in (12), considering data transfers of duration $\gamma D$. Conversely, the result derived in (7) does not accounts properly for the impact of half-duplex connections whose duration is different from the one of the information unit being decoded, and needs to be extended. A similar reasoning applies to the success probability of a half-duplex link, making it possible to rely on the Laplace transform of the interference generated by clusters of the same kind in (7) while requiring a new computation for the impact of full-duplex communications with different duration. In summary, we have

$$p_s^{(\mathsf{hd})} = \mathcal{L}_{\mathcal{I}_{\mathsf{hd}}}\big(\theta r^\alpha, D\big)\, \mathcal{L}_{\mathcal{I}_{\mathsf{fd,hd}}}\big(\theta r^\alpha\big) \tag{19}$$

$$p_s^{(\mathsf{fd})} = \mathcal{L}_{\mathcal{I}_{\mathsf{fd}}}\big(\theta r^\alpha, \gamma D\big)\, \mathcal{L}_{\mathcal{I}_{\mathsf{hd,fd}}}\big(\theta r^\alpha\big) \cdot \beta \tag{20}$$

where the first factors in (19) and (20) are the Laplace transforms derived in Section III with an additional argument specifying the packet duration that shall be accounted for in the corresponding exponential function; $\mathcal{L}_{\mathcal{I}_{\mathsf{fd,hd}}}$ accounts for the interference generated by full-duplex links of duration $\gamma D$ over a half-duplex reception; and $\mathcal{L}_{\mathcal{I}_{\mathsf{hd,fd}}}$ covers the interference affecting a full-duplex receiver due to half-duplex clusters transmitting for a time $D$.

Let us initially focus on the last term. By definition, and following the same steps discussed for the derivation of (10), we can write for $s \geq 0$

$$\begin{aligned} \mathcal{L}_{\mathcal{I}_{\mathsf{hd,fd}}}(s) &= \mathbb{E}\Big[ \prod_{(\mathbf{u},T)\in\Lambda_{\mathsf{hd}}} e^{-s\omega'(T,\gamma)L(\mathbf{u})\zeta_{\mathbf{u}}} \Big] \\ &= \exp\left( -\lambda(1-q) \int_{\mathbb{R}^2} d\mathbf{u} \int_{\mathbb{R}} \left( 1 - \frac{1}{1+sL(\mathbf{u})\omega'(T,\gamma)} \right) dT \right) \end{aligned} \tag{21}$$

where the ancillary function reporting for the fraction of the half-duplex transmission of cluster $(\mathbf{u}, T)$ that overlaps with the bidirectional link of interest is slightly reformulated as

$$\omega'(T,\gamma) := \frac{1}{\gamma D}\int_0^{\gamma D} \mathbb{I}(T \le t \le T + D)\, dt.$$

If we concentrate for the moment on the scenario $\gamma \le 1$, simple calculations show that $\omega'$ has support $[-D, \gamma D]$ and evaluates within it to

$$\omega'(T,\gamma \le 1) = \frac{1}{\gamma D}\cdot \begin{cases} D + T & T \in [-D, -D(1-\gamma)) \\ \gamma D & T \in [-D(1-\gamma), 0) \\ \gamma D - T & T \in [0, \gamma D] \end{cases}$$

This outcome allows to explicitly solve the time integral in (21), which can be expressed as

$$2\gamma D\left(1 - \frac{\ln(1+sL(\mathbf{u}))}{sL(\mathbf{u})} + \frac{1-\gamma}{2\gamma}\left(1 - \frac{1}{1+sL(\mathbf{u})}\right)\right) \quad (22)$$

Finally, plugging (22) into (21) and observing that the integration over the spatial component is independent of the angular coordinate of the interfering half-duplex cluster, it is possible to derive a closed form expression for the sought Laplace transform:

$$\mathcal{L}_{\mathcal{I}_{\mathsf{hd,fd}}}(\theta r^{\alpha} | \gamma \le 1) = \exp\Big(-\lambda(1-q)\gamma D\, \Omega'_{\mathsf{hd}}\Big) \quad (23)$$

resorting once more to the auxiliary function

$$\begin{aligned}\Omega'_{\mathsf{hd}}&(r,\theta,\alpha\,|\,\gamma \le 1) \\ &= \pi r^2 \theta^{\frac{2}{\alpha}} \Gamma\left(1 + \frac{2}{\alpha}\right)\Gamma\left(1 - \frac{2}{\alpha}\right)\left(\frac{2\alpha}{2+\alpha} + \frac{2(1-\gamma)}{\gamma}\right).\end{aligned} \quad (24)$$

The achieved result is quite insightful, as it provides a neat extension of the reference case studied in Section III. Firstly, comparing (23) to the Laplace transform of half-duplex interference reported in (7), we infer that the role played by the packet duration $D$ and the traffic density $\lambda$ is decoupled from the the other system parameters even when links of different durations are allowed in the network. Even more interestingly, (24) captures the effect of longer half-duplex communications over a full-duplex data exchange with respect to its counterpart in (8) simply by means of an additive correction term embodied by $2(1-\gamma)/\gamma$.

To complete the performance characterisation in terms of success probability, the evaluation of the impact of interference generated by full-duplex pairs – quantified by $\mathcal{L}_{\mathcal{I}_{\mathsf{fd,hd}}}$ – is in order. Considering again the case $\gamma \le 1$, the definition of Laplace transform leads us to

$$\mathcal{L}_{\mathcal{I}_{\mathsf{fd,hd}}}(s) = \mathbb{E}\left[\prod_{(\mathbf{u},T)\in\Lambda_{\mathsf{fd}}} e^{-s\omega''(T,\gamma)(L(\mathbf{u})\zeta_{\mathbf{u}} + L(\mathbf{v})\zeta_{\mathbf{v}})}\right]$$

where $s \in \mathbb{R}^{+}$ and the auxiliary function $\omega''$ expresses the average fraction of the half-duplex reception interfered by the full-duplex transmissions in cluster $(\mathbf{u}, T)$ as

$$\omega''(T,\gamma) := \frac{1}{D}\int_0^{D} \mathbb{I}(T \le t \le T + \gamma D)\, dt\,. \quad (25)$$

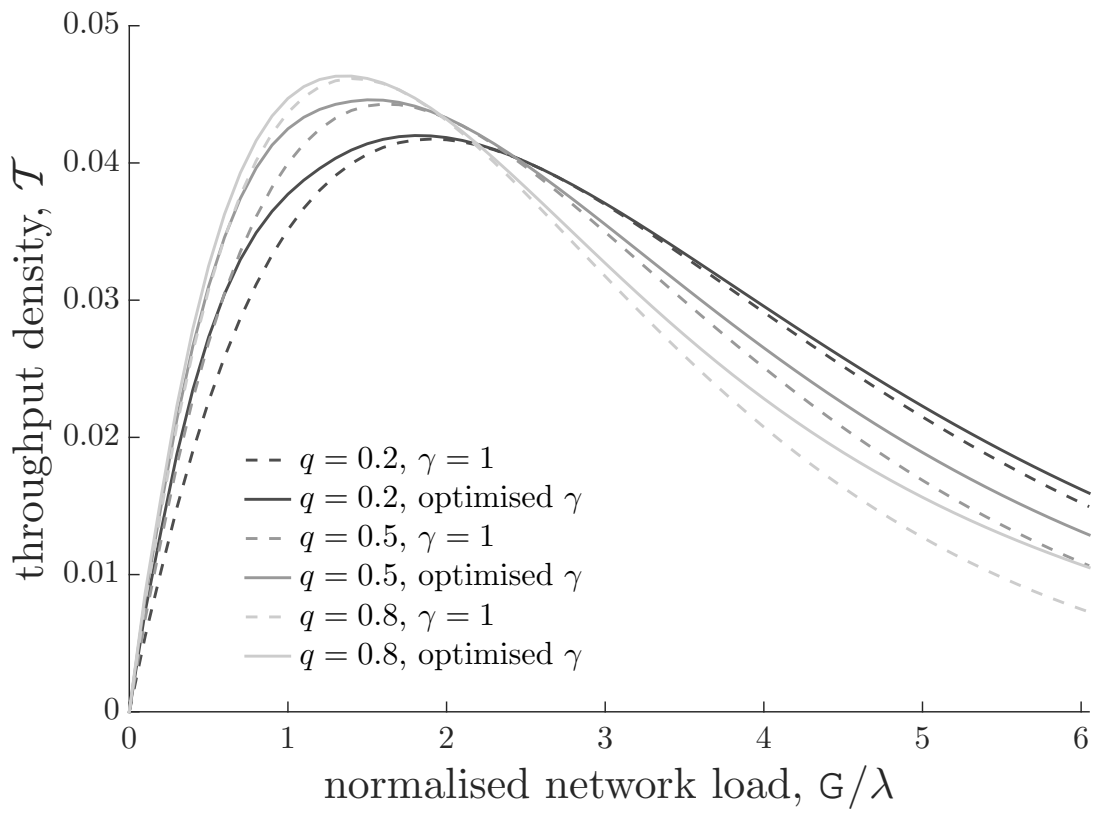

Fig. 8: Throughput density vs normalised network load $\mathsf{G}/\lambda$. Dashed lines report the behaviour of a homogeneous system with all transmissions of the same duration (i.e. $\gamma = 1$), whereas solid ones indicate the performance achieved when optimising $\gamma$ in the heterogeneous case. Different shades of grey indicate distinct values of $q$.

(25) paves the road for the mathematical derivation of $\mathcal{L}_{\mathcal{I}_{\mathsf{fd,hd}}}$, which proceeds along the same footsteps taken in Section III. While conceptually similar, the involved calculations are rather cumbersome, all the more so if we observe that for $\gamma > 1$ slightly different structures are obtained for the $\omega'$ and $\omega''$ functions, prompting further integrations to be tackled. For the sake of compactness we thus omit the details of the derivation, and report the key result, eventually expressing the success probabilities in a network with half- and full-duplex links of duration $D$ and $\gamma D$ respectively as

$$\begin{aligned} p_s^{(\mathsf{hd})} &= \exp\Big(-\lambda D\Big((1-q)\,\Omega_{\mathsf{hd}} + q\,\Omega'_{\mathsf{fd}}\Big)\Big) \\ p_s^{(\mathsf{fd})} &= \beta\cdot\exp\Big(-\lambda\gamma D\Big((1-q)\,\Omega'_{\mathsf{hd}} + q\,\Omega_{\mathsf{fd}}\Big)\Big). \end{aligned}$$

Here, $\Omega_{\mathsf{hd}}$ and $\Omega_{\mathsf{fd}}$ are the functions already discussed in (8) and (13), whereas $\Omega'_{\mathsf{hd}}$ and $\Omega'_{\mathsf{fd}}$ are summarised for any value of $\gamma$ in equations (27)-(28) at the bottom of next page. For completeness, we also include in (26) the explicit structure of the ancillary functions $\omega'(t,\gamma)$ and $\omega''(t,\gamma)$ which are used to solve the integrals leading to the Laplace transforms.

The presented framework extension enables thus a direct comparison with the *homogeneous-duration* setting discussed in the initial part of the paper. A first question of interest is whether and how much one could gain in a *heterogeneous-duration* case by letting half- and full-duplex connections have different durations $D_{\mathsf{hd}}$ and $D_{\mathsf{fd}} = \gamma D_{\mathsf{hd}}$. To gather a sound answer, we compare the two system configurations under the same channel occupancy conditions, i.e. $\mathsf{G} = \lambda D$ in the homogeneous case and $\mathsf{G} = \lambda D_{\mathsf{hd}}(1 + q(\gamma - 1))$ in the heterogeneous case. Moreover, fixing the traffic intensity $\lambda$, for any value of $\mathsf{G}$ we operate the heterogeneous network under the $(D_{\mathsf{hd}}, \gamma)$ pair maximising the throughput density in (18), so to understand what is the utmost improvement that can be aimed for. The outcome of this study is reported in Fig. 8 in terms of $\mathcal{T}$ against the normalised load $\mathsf{G}/\lambda$.[9] The reference parameters $\lambda$, $r$, $\theta$, $\alpha$ have been kept as in Section III, and perfect self-interference cancellation is assumed. Within the plot, dashed lines indicate the behaviour of the homogeneous system, whereas solid ones mark the performance of the optimised heterogeneous configuration. Moreover, three sets of curves are reported in different shades of grey, referring to distinct fractions of full-duplex clusters present in the network.

The figure confirms the intuition that operating a hybrid system with a common duration for all communications is not optimal when a completely asynchronous medium access policy is employed. As expected, an heterogeneous setup is especially beneficial when the network is either experiencing low loads or facing congestion. In the former situation, in fact, throughput can be boosted by granting longer data exchanges to bidirectional links while shortening the less profitable half-duplex clusters. Conversely, when high loads are experienced, a reduction in the duration of full-duplex communications in favour of uni-directional ones maps into a lower level of aggregate interference and thus induces a more gentle degradation of the performance. As a result, the additional degree of freedom granted to the system can triggering throughput gains in the order of 15-20% for a broad range of $q$ values. On the other hand, such an improvement comes at the cost of a potential unfairness among users, possibly constraining some of them to access the medium only for short information exchanges.

From this standpoint it is thus also insightful to consider the complementary scenario of an existing network characterised by a population of half-duplex clusters transmitting packets of duration $D_{\mathsf{hd}}$, and investigate how to tune data transfers for advanced full-duplex capable nodes that are progressively introduced into the system. More formally, we are interested in determining the duration ratio $\gamma^*$ maximising the aggregate throughput density for a certain penetration level $q$. The solution can be found via numerical optimisation of (18), and is shown in Fig. 9, where distinct lines indicate different configurations in terms of the half-duplex packet duration. For large values of $D_{\mathsf{hd}}$, the reference network is already operating with a medium to heavily congested channel, and new full-duplex links shall enjoy much shorter communications not to increase too much the overall interference, regardless of $q$. On the other hand, when smaller data units are employed by half-duplex pairs (e.g. $D_{\mathsf{hd}} \leq 1$), a notable fraction of the clusters can be upgraded to bidirectional mode and granted channel access for longer fraction of time ($\gamma^* > 1$) bringing an improvement to the system throughput. In this perspective, the developed framework offers then useful tools to evaluate whether it is worth to undergo the cost of deploying full-duplex terminals given the working conditions of an existing topology, taking into account the traffic profiles and applications such new nodes shall be able to sustain.

## V. A Comparison with Slotted Schemes

We conclude our study by investigating the performance gap between the asynchronous network under consideration and a slotted counterpart of its. The rationale triggering such a discussion is twofold. On the one hand, the tradeoff

[9] The normalised load $\mathsf{G}/\lambda$ maps to the packet duration $D$ in homogeneous-duration networks, so that the $x$-axis in the plot is conveniently equivalent to the ones characterising figures discussed in the first part of the paper.

$$\omega'(T,\gamma\leq 1)=\frac{1}{\gamma D}\cdot\begin{cases}D+T & T\in[-D,-D(1-\gamma))\\ \gamma D & T\in[-D(1-\gamma),0)\\ \gamma D-T & T\in[0,\gamma D]\end{cases}\qquad \omega'(T,\gamma> 1)=\frac{1}{\gamma D}\cdot\begin{cases}D+T & T\in[-D,0)\\ D & T\in[0,D(\gamma-1))\\ \gamma D-T & T\in[D(\gamma-1)),\gamma D]\end{cases}$$

$$\omega''(T,\gamma\leq 1)=\frac{1}{D}\cdot\begin{cases}\gamma D+T & T\in[-\gamma D,0)\\ \gamma D & T\in[0,D(\gamma-1))\\ \gamma D-T & T\in[D(\gamma-1),D]\end{cases}\qquad \omega''(T,\gamma> 1)=\frac{1}{D}\cdot\begin{cases}\gamma D+T & T\in[-D\gamma,-D(\gamma-1))\\ D & T\in[-D(\gamma-1),0)\\ \gamma D-T & T\in[0,D]\end{cases} \tag{26}$$

$$\begin{aligned}\Omega'_{\mathsf{hd}}(r,\theta,\alpha\,|\,\gamma\leq 1)&=\pi r^2\theta^{\frac{2}{\alpha}}\Gamma\left(1+\frac{2}{\alpha}\right)\Gamma\left(1-\frac{2}{\alpha}\right)\left(\frac{2\alpha}{2+\alpha}+\frac{2(1-\gamma)}{\gamma}\right)\\ \Omega'_{\mathsf{hd}}(r,\theta,\alpha\,|\,\gamma> 1)&=\pi r^2\theta^{\frac{2}{\alpha}}\Gamma\left(1+\frac{2}{\alpha}\right)\Gamma\left(1-\frac{2}{\alpha}\right)\left(\frac{2\alpha}{2+\alpha}+\frac{\gamma-1}{\gamma}\right)\gamma^{-(1+2/\alpha)}\end{aligned} \tag{27}$$

$$\begin{aligned}\Omega'_{\mathsf{fd}}(r,\theta,\alpha\,|\,\gamma\leq 1)&=\int_0^\infty 4u\left(\frac{\pi(1+\gamma)}{2}-\int_0^\pi \frac{\ln\left(\frac{1+\gamma\theta r^\alpha u^{-\alpha}}{1+\gamma\theta r^\alpha \ell(u,\varphi)}\right)}{\theta r^\alpha\left(u^{-\alpha}-\ell(u,\varphi)\right)}-\frac{1-\gamma}{2(1+\gamma\theta r^\alpha u^{-\alpha})(1+\gamma\theta r^\alpha\ell(u,\varphi))}\,d\varphi\right)du\\ \Omega'_{\mathsf{fd}}(r,\theta,\alpha\,|\,\gamma> 1)&=\int_0^\infty 4u\left(\frac{\pi(1+\gamma)}{2}-\int_0^\pi \frac{\ln\left(\frac{1+\theta r^\alpha u^{-\alpha}}{1+\theta r^\alpha \ell(u,\varphi)}\right)}{\theta r^\alpha\left(u^{-\alpha}-\ell(u,\varphi)\right)}-\frac{\gamma-1}{2(1+\theta r^\alpha u^{-\alpha})(1+\theta r^\alpha\ell(u,\varphi))}\,d\varphi\right)du\end{aligned} \tag{28}$$

where, $\ell(u,\varphi)=\left(u^2+r^2+2ru\cos\varphi\right)^{-\frac{\alpha}{2}}$

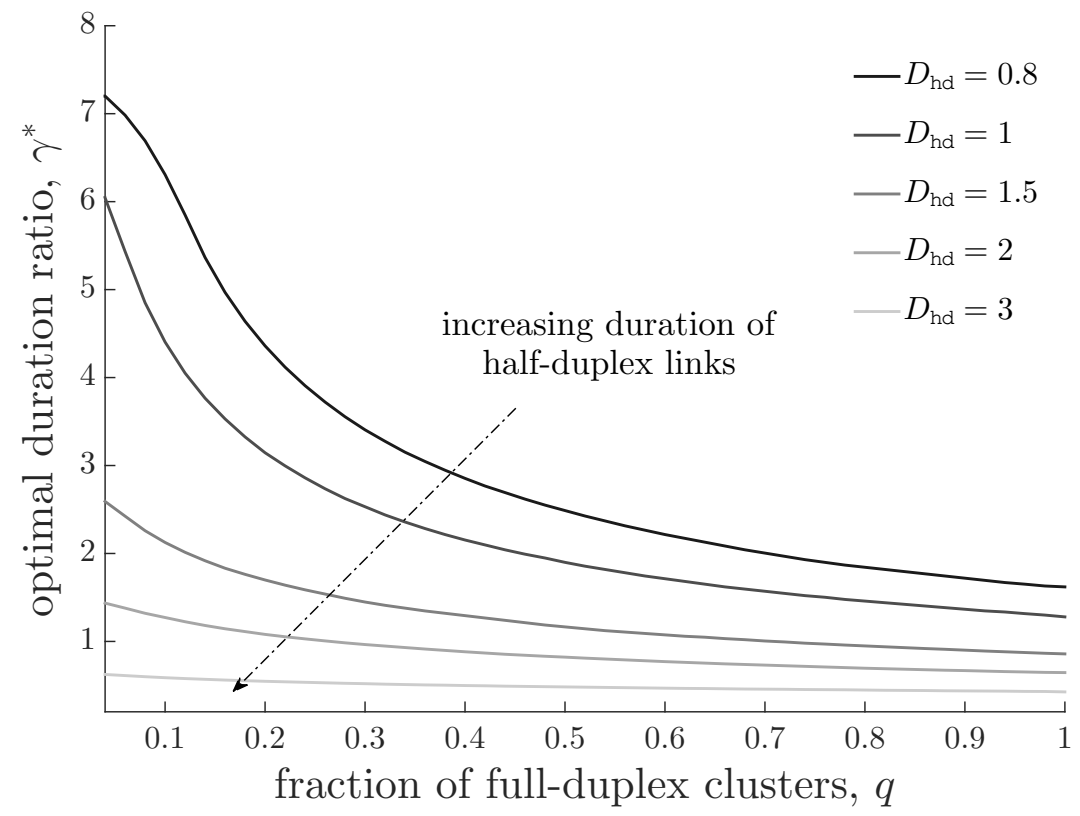

Fig. 9: Ratio of full-duplex to half-duplex transmission duration maximising throughput as a function of $q$. Different shades of grey report the behaviour for distinct durations $D_{\mathsf{hd}}$ of the half-duplex links.

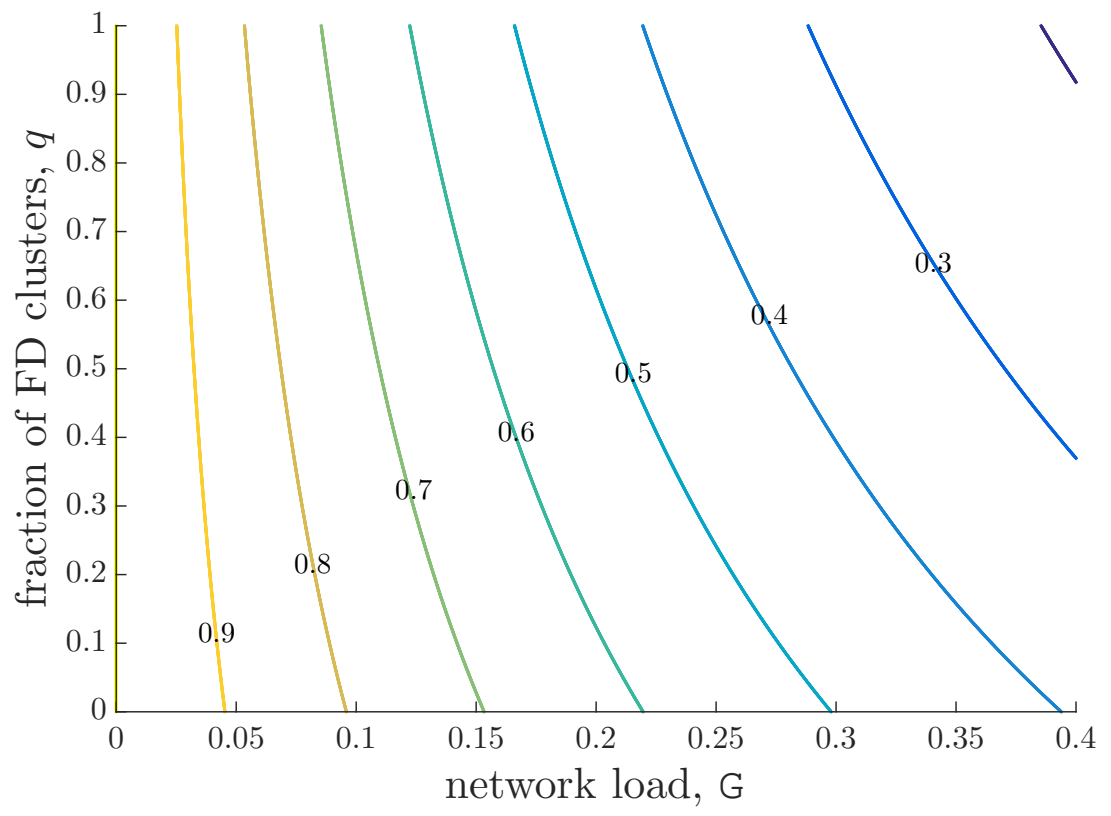

Fig. 10: Ratio $\Xi$ of the throughput density of an unslotted system over a slotted one as a function of the fraction of full-duplex clusters in the network and the network load. All transmissions in the asynchronous system are of common duration.

between the throughput loss undergone by systems that do not implement any form of coordination among terminals and their intrinsic simplicity is often a basic driver for protocol design and implementation choices and, as such, has been tackled starting from the seminal works of Abramson on Aloha [31]. From this viewpoint, while recent results based on stochastic geometry [25], [30] have provided interesting insights for large and distributed networks, no characterisation is available yet for systems that resort to full-duplex communications. Secondly, the study will allow us to understand the impact of the additional degree of freedom in terms of different packet durations for half- and full-duplex clusters available in unslotted systems and to clarify whether it can help in reducing the performance degradation with respect to synchronous ones. Throughout our discussion, we will refer to the elegant analysis of slotted full-duplex networks offered in [18], and point the interested reader to it for further details. In the following, we recall the key results needed for our comparison, and highlight the main conceptual differences with respect to the framework introduced in Section II. In the synchronous scenario, the topology is still composed of pairs of nodes that independently decide whether to establish a bidirectional or a half-duplex link with probability $q$ and $1-q$, respectively. Time is divided in slots of equal duration, and each data exchange in the system can be performed only within the boundaries of – and fill completely – one such time unit. This element of coordination allows for some major analytical simplifications. Firstly, it decouples the time and spatial components of medium access, so that the network can be effectively described by means of a PPP over $\mathbb{R}^2$ of intensity $\lambda_s$, with each node pair independently deciding whether to access a slot for an information transfer with probability $p$. Secondly, the shared time-frame results in a constant level of interference perceived at a receiver for the whole duration of an incoming packet, allowing to overlook the averaging procedures tackled in our work. To ensure a fair comparison, decoding is described via a threshold model for the slotted system as well, and all networking parameters are unchanged with respect to the presented framework. Lastly, we are still interested in evaluating the behaviour of the system as a function of the load, which we defined as the fraction of time the channel is occupied on average in a unit area.

When synchronous access is considered, the dependency of this parameter on the duration of transmissions is clearly lost, leading to $\mathsf{G} = p\lambda_s$. Under these modelling assumptions, the system throughput density can be eventually expressed as [18]

$$\mathcal{T}_s = W\mathsf{G}\big(1+q(2\beta-1)\big)\cdot e^{-\mathsf{G}\left((1-q)\Omega_{\mathsf{hd},s}+q\,\Omega_{\mathsf{fd},s}\right)} \tag{29}$$

where

$$\Omega_{\mathsf{hd},s} = \pi r^2\theta^{\frac{2}{\alpha}}\Gamma\left(1+\frac{2}{\alpha}\right)\Gamma\left(1-\frac{2}{\alpha}\right)$$
$$\Omega_{\mathsf{fd},s} = \int_0^\infty 2u\Big(\pi-\frac{1}{1+\theta r^\alpha u^{-\alpha}}\int_0^\pi \frac{1}{1+\theta r^\alpha \ell(u,\varphi)}d\varphi\Big)du$$

and $\ell(u,\varphi)$ is defined in (28).

As anticipated in the previous sections, the ancillary $\Omega$ function that captures the effect of full-duplex interference in a slotted system is structurally distinct from the one derived for asynchronous Aloha, driving the different achievable performance. To better understand and quantify this aspect, we focus on the metric $\Xi$, defined as the ratio of the throughput of a completely asynchronous configuration to the one of its synchronous counterpart operated at the same network load $\mathsf{G}$, with equal parameters and fraction of full-duplex clusters. Let us initially assume for the unslotted network all data transmissions to be of the same duration. In this case, a direct comparison of the throughput expressions in (14) and (29) shows how imperfect self-interference cancellation besets the two scenarios in the same way, so that $\Xi$ is in fact independent of $\beta$ and captures the intrinsic differences between the two access policies beyond specific implementation aspects. The behaviour of the throughput ratio is reported in Fig. 10 as a function of $\mathsf{G}$ and of the penetration level $q$ of full-duplex pairs, highlighting how the performance gap widens when the network faces larger network loads, and confirming the intuition that an increase of aggregate interference is more detrimental in a fully uncoordinated scenario.[10] This rationale also buttresses the similar yet less pronounced trend that can be spotted when more full-duplex transmissions are triggered.

In this perspective, additional insights are offered by Fig. 11, which reports $\Xi$ against $q$ for three different load configurations. Let us first focus on the solid lines, representative of the behaviour of an asynchronous system with common duration for all transmissions, and consider in particular the values for $q = 0$. Such points indicate the performance loss brought by the lack of a slotted time-frame in a traditional completely half-duplex network. Remarkably, when light loads are tackled, synchronism among nodes throughout the network only triggers a 10% gain, making unslotted access particularly attractive in view of its simplicity. Conversely, under strong congestion (e.g. $\mathsf{G} = 0.35$), the performance of an asynchronous MAC plummets to less than half of its slotted competitor. The plot also sheds light on the impact of the additional interference brought by full-duplex connections. It is interesting to observe in fact that, while spatial reuse is better taken advantage of in slotted systems, the throughput loss undergone in the asynchronous case when increasing $q$ is rather contained, especially for low-to-intermediate load. Along this line of reasoning, it is then relevant to understand whether the gap may be further reduced by leveraging the additional degree of freedom of different transmission durations available in unslotted settings. The question is tackled once more in Fig. 11, where dashed lines report, for any value of $q$, the performance degradation $\Xi$ undergone by an heterogeneous asynchronous setting whose parameter pair $(\gamma, D)$ has been configured so as to optimise the throughput in (18). It is apparent how a smarter subdivision of resources in terms of channel occupation time among half- and full-duplex links can partly counterbalance the inefficiency induced by uncoordinated transmissions. Such a result becomes especially remarkable since the benefits are attainable in operating regions of practical interest. For example, if we recall that the traffic intensity of the asynchronous network has been set to $\lambda = 0.05$ throughout our discussion, we can infer that the dashed curve for $\mathsf{G} = 0.05$ in Fig. 11 describes the behaviour of the system for a normalised network load $\mathsf{G}/\lambda = 1$. This value, in turn, is shown in Fig. 8 to offer a throughput density close to its peak and is thus representative of a working point typically targeted for efficient network operations.

Beyond numerical results, however, the analytical comparison of slotted and unslotted full-duplex systems enabled by the developed framework shall be seen as a tool towards an educated choice on whether to strive for synchronism or not when designing a system, as it offers insights on some of the key involved performance tradeoffs.

## VI. Conclusions

This paper introduced a stochastic geometry framework to capture the performance of an asynchronous Aloha network where part of the nodes operate in full-duplex mode. Exact expressions have been presented for the success probability and the system throughput, prompting the key tradeoffs in the system. In particular, three operating regions have been identified, showing how for short enough packets as many communications as possible shall be performed in full-duplex mode, while for packets longer than a certain threshold solely relying on half-duplex is convenient. Under the assumption of complete self-interference cancellation, the maximum throughput gain achievable over a purely half-duplex system has been proven to be independent of the distance between source and destination of a link. Bringing imperfect cancellation into the picture, instead, the distance between two communicating nodes becomes a critical parameter, with full-duplex paying off only over short links. An optimisation approach leveraging different link durations for bidirectional and unidirectional data exchanges was introduced to improve network performance and, finally, a comparison with the slotted case studied in [18] was discussed, clarifying the effectiveness cost undergone for not synchronising medium access among devices. In all settings, the role of very accurate self-interference cancellation schemes has been confirmed as a necessary condition for full-duplex to be convenient in broad and uncoordinated networking scenarios.

[10]For the sake of comparison, note for instance that for the considered $\lambda = 0.05$, a network load $\mathsf{G} = 0.2$ in Fig 10 corresponds to a packet duration $D = 4$ in the plots of Section III (e.g. Fig. 3).

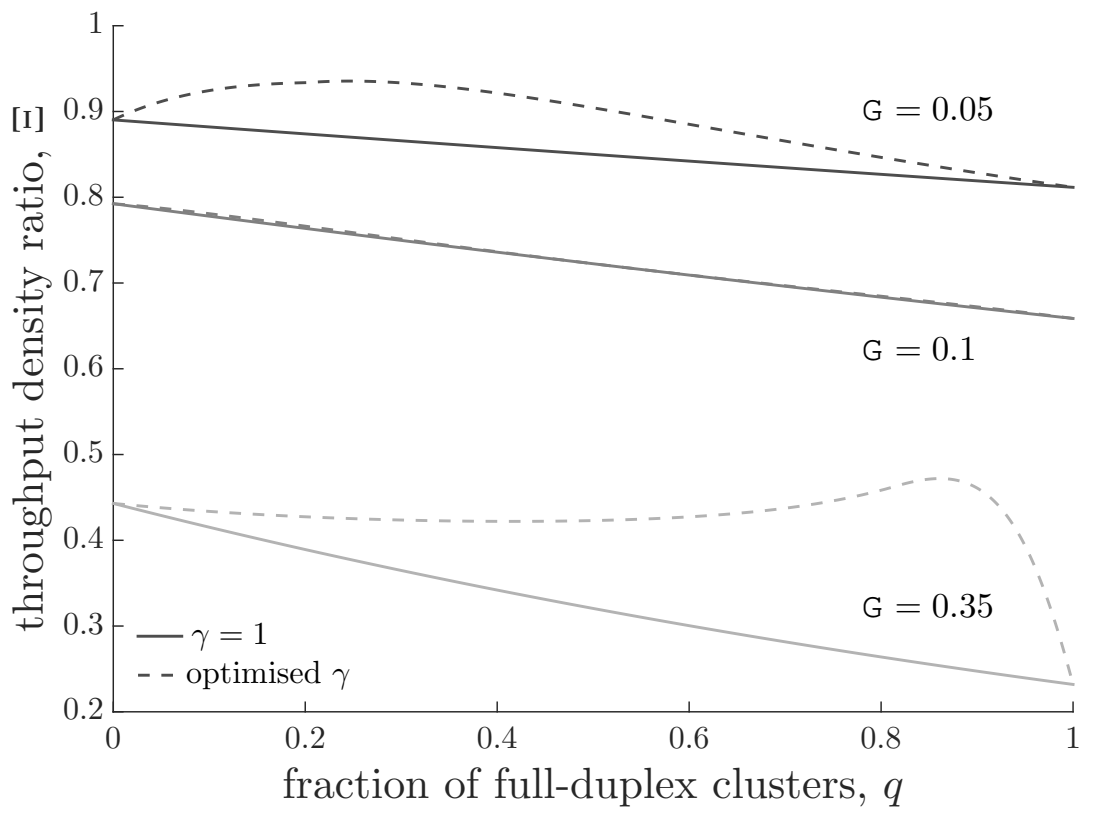

Fig. 11: Ratio $\Xi$ of the throughput density of an unslotted system over a slotted one vs $q$. Solid lines report the ratio when the asynchronous system is operated with all transmissions of the same duration, whereas dashed ones consider optimal values of $\gamma$. Different shades of grey indicate distinct network loads.

## Appendix A
## Proof of Theorem 1

We aim to show that $\Omega_{\mathsf{fd}}(r, \theta, \alpha)/\Omega_{\mathsf{hd}}(r, \theta, \alpha)$ does not depend on the distance $r$ between the two nodes in a cluster. Recalling from (8) that the denominator can be written in the form $\Omega_{\mathsf{hd}}(r, \theta, \alpha) = k(\theta, \alpha) r^2$, $k \in \mathbb{R}$, the proposition is proven as soon as the numerator exhibits a quadratic dependence on $r$ as well. As a first step, we observe that the definition of $\Omega_{\mathsf{fd}}$ in (13) leverages symmetry to compute the spatial average over the cluster centre considering a node $\mathbf{u}$ moving along the $x$ axis, i.e. $\mathbf{u} = u e^{j\xi}$, $\xi = 0$. Leaning on the expression $\mathbf{w} = r e^{j\varphi}$ and on the notation of Fig. 1, we can thus reformulate (13) as $\Omega_{\mathsf{fd}} = \int_0^\infty 4u g(u)\, du$, where

$$g(u) = \int_0^\pi 1 - \frac{\ln\left(1+sL(\mathbf{u})\right) - \ln\left(1+sL(\mathbf{u}+\mathbf{w})\right)}{s\left(L(\mathbf{u}) - L(\mathbf{u}+\mathbf{w})\right)}\, d\varphi$$

and $s = \theta r^\alpha$. For a given $\mathbf{u}$ on the $x$-axis, the integral solely depends on the path loss function $L(\cdot)$ computed at the companion node in the cluster, which in turn is maximised for $\varphi = \pi$, i.e. when $\mathbf{w} = \mathbf{w}' = -r$. For the integrand within $g(u)$ we hence obtain

$$\begin{aligned} &1 - \frac{\ln\left(1+sL(\mathbf{u})\right) - \ln\left(1+sL(\mathbf{u}+\mathbf{w})\right)}{s\left(L(\mathbf{u}) - L(\mathbf{u}+\mathbf{w})\right)} \\ &\overset{(a)}{\leq} 1 - \frac{\ln\left(\frac{1+su^{-\alpha}}{1+s|u-r|^{-\alpha}}\right)}{s\left(u^{-\alpha} - |u-r|^{-\alpha}\right)} \overset{(b)}{\leq} 1 - \frac{1}{1+su^{-\alpha}}. \end{aligned} \tag{30}$$

The first inequality upper bounds the denominator as $L(\mathbf{u}) - L(\mathbf{w}) \leq L(\mathbf{u}) - L(\mathbf{w}')$ and lower bounds the numerator as

$$\ln\left(\frac{1+sL(\mathbf{u})}{1+sL(\mathbf{u}+\mathbf{w})}\right) \geq \ln\left(\frac{1+sL(\mathbf{u})}{1+sL(\mathbf{u}+\mathbf{w}')}\right)$$

for any $\varphi \in [0,\pi]$. Conversely, inequality (b) follows by applying to the logarithmic numerator the well-known relation $\ln(x) \geq 1 - 1/x$, $x \in \mathbb{R}$ and by carrying out simple manipulations on the obtained expression. Since the rightmost expression in (30) is independent of $\varphi$, we get

$$g(u) \leq \pi\left(1 - \frac{1}{1+su^{-\alpha}}\right).$$

Plugging this into the definition of $\Omega_{\mathsf{fd}}$, and evaluating the integral over $u$, we eventually get

$$\Omega_{\mathsf{fd}} \leq 4r^2\left(\pi\theta^{\frac{2}{\alpha}}\, \Gamma(1+2/\alpha)\, \Gamma(1-2/\alpha)\, \alpha\right). \tag{31}$$

To complement this result, let us focus on the two cases of a solely half-duplex and a solely full-duplex network with ideal self-interference cancellation, corresponding to $q = 0$ and $q = 1$, respectively. Assuming the same set of parameters for the two scenarios, particularly in terms of density $\lambda$ and link duration $D$, $p_s^{(\mathsf{hd})} \geq p_s^{(\mathsf{fd})}$ clearly holds, as the full-duplex system undergoes on average a larger level of interference due to the increased spatial reuse. Recalling (4)-(5) and the expressions of the Laplace transforms in (7)-(8), (12), we then get $\exp(-\lambda D \Omega_{\mathsf{hd}}) \geq \exp(-\lambda D \Omega_{\mathsf{fd}})$, leading to

$$\Omega_{\mathsf{fd}} \geq r^2\left(\pi\theta^{\frac{2}{\alpha}}\, \Gamma(1+2/\alpha)\, \Gamma(1-2/\alpha)\, \frac{2\alpha}{\alpha+2}\right). \tag{32}$$

Combining (31) and (32), the real-valued analytical function $\Omega_{\mathsf{fd}}(r,\theta,\alpha)$ is lower- and upper-bounded for any $r \in \mathbb{R}$ by curves in the form $Ar^2$ and $Br^2$, with $A$ and $B$ real constants. The statement under proof then readily follows from elementary applications of analytical geometry [32]. ∎